\documentclass[preprint,12pt]{elsarticle}

\usepackage{amssymb}
\usepackage{natbib}
\usepackage{amsmath}
\usepackage{xcolor}
\usepackage{siunitx}
\usepackage{array}
\usepackage{geometry}
\usepackage{booktabs}
\begin{document}

\begin{frontmatter}

\title{Validation of the GFS model for Gyrokinetic Stability of NSTX Pedestal Data} 

\author[1]{M. Yang} %% Author name
\author[2]{J.F. Parisi}
\author[1]{M. S. Anastopoulos Tzanis}
\author[1]{G.M. Staebler}

\affiliation[1]{organization={Oak Ridge National Laboratory}, 
            city={Oak Ridge},
            state={Tennessee},
            country={USA}}
\affiliation[2]{organization={Princeton Plasma Physics Laboratory}, 
            city={Princeton},
            state={New Jersey},
            country={USA}}
%% Abstract
\begin{abstract}
% This study presents a large database verification of the Gyro Fluid System (GFS) model for linear gyrokinetic stability for high-mode (H-mode) edge transport barrier conditions in the National Spherical Torus Experiment (NSTX) tokamak. The database of linear stability calculations with the CGYRO gyrokinetic code was produced using plasma profile measurements from NSTX discharges to identify the most unstable modes that limit the pressure profile gradient in the H-mode barrier in a previous study. A novel Bayesian optimization approach determines optimal resolution parameters for GFS specifically tailored for spherical tokamak pedestal conditions. Our results demonstrate that GFS with low velocity space resolution can achieve accurate linear stability analysis in NSTX pedestal conditions. At high-resolution GFS achieves comparable growth rate accuracy but exhibits a few percent high frequency outliers in the lower magnetic shear region near the top of the H-mode pedestal. These findings establish GFS as a fast linear eigenmode solver for spherical tokamak pedestal gyrokinetic stability and demonstrate a systematic methodology for determining the optimum resolution settings.

This study presents a large database validation of the Gyro Fluid System (GFS) model for linear gyrokinetic stability for high-mode (H-mode) edge transport barrier conditions in the National Spherical Torus Experiment (NSTX) tokamak. The database of linear stability calculations with the CGYRO gyrokinetic code was produced using plasma profile measurements from NSTX discharges to identify kinetic ballooning modes (KBM), trapped electron modes (TEM), and micro-tearing modes (MTM) that limit the pressure profile gradient in the H-mode barrier. A novel Bayesian optimization approach determines optimal resolution parameters for GFS specifically  for spherical tokamak pedestal conditions. Our results demonstrate that GFS, with optimized resolution, can achieve accurate linear stability analysis in NSTX pedestal conditions for reduced resolution compared to CGYRO. GFS can accurately find the KBM, TEM, and MTM instability branches. Parametric analysis reveals that GFS accuracy in this extreme pedestal parameter range is degraded for low magnetic shear and near the separatrix conditions. These findings establish GFS as a fast linear eigenmode solver for spherical tokamak pedestal gyrokinetic stability and demonstrate a systematic methodology for determining the optimum resolution settings.
\end{abstract}

%% Keywords

\end{frontmatter}

\section{Introduction}
\label{sec:intro}

Understanding plasma transport is crucial for optimizing magnetic confinement fusion devices and ensuring reliable transport predictions for current and future fusion reactors. While gyrokinetic turbulence simulations can simulate plasma transport, they demand substantial computational resources that become impractical for electron-ion multi-scale simulations and time dependent plasma profile prediction. To design next-generation fusion machines, reduced transport models must be validated across different plasma conditions, aspect ratios and magnetic fields to determine optimal operational parameters. Reduced quasi-linear transport models like the Trapped Gyro-Landau Fluid (TGLF) model \cite{staebler2007theory} are able to reproduce the turbulent fluxes from gyrokinetic simulations to within 20\% for core plasma conditions in conventional aspect ratio tokamaks \cite{Staebler:2021}. However, at higher  plasma pressure/magnetic pressure ($\beta$), the TGLF linear eigenmode solver is not very accurate at finding the threshold for the kinetic ballooning mode (KBM) instability, especially at low aspect ratio \cite{kinsey2025analysis}, even for core plasma conditions. A new gyro-fluid system (GFS) \cite{staebler2023flexible} of linear moment equations was developed to allow for a flexible velocity space resolution to achieve higher accuracy than TGLF while retaining a speed advantage over other gyrokinetic linear solvers due to its fully spectral numerical method and novel closure. The GFS linear growthrates have been shown to be more accurate than TGLF for core tokamak \cite{staebler2023flexible} and spherical tokamak \cite{kinsey2025analysis} conditions particularly at high $\beta$. 

The extreme plasma gradients, strong magnetic shear and high safety factor of the H-mode pedestal are challenging for linear Magnetohydrodynamic (MHD) and gyrokinetic stability calculations. The EPED model is able to predict the width and height of a hyperbolic tangent trial profile for the plasma pressure by a combination of the linear MHD stability constraints for peeling-ballooning (PBM) and ideal ballooning modes (IBM) \cite{snyder2009pedestal}. The global plasma energy can be predicted in a tokamak using the EPED model for the pedestal as a boundary condition for a core plasma transport solution \cite{park:2018b}. 

Recent research \cite{Dickinson2012,parisi2024kinetic,Parisi_2024b,Parisi_2024c} has identified KBM, rather than IBM instabilities, as the primary mechanism limiting plasma confinement in spherical tokamak pedestals. Linear gyrokinetic analysis revealed that KBMs become unstable at roughly half the IBM threshold, explaining the distinctive pedestal width scaling observed in NSTX \cite{Diallo2013, snyder2009pedestal}. This finding has significant implications for fusion power optimization in spherical tokamaks, since performance correlates with pedestal pressure \cite{menard2016fusion}. High-$\beta$ operation at low aspect ratio introduces electromagnetic effects that can destabilize both Micro Tearing Modes (MTMs) and KBMs in the plasma pedestal and core regions.

Spherical tokamaks like NSTX \cite{nstx} present distinct challenges compared to conventional aspect-ratio devices. The compact geometry and broader edge pedestals characteristic of low aspect-ratio configurations \cite{Berkery2024,Harrison2024} require an EPED type model to use KBM rather than IBM to predict the observed pedestal width. The first goal of our study is to assess the ability of GFS to provide the KBM constraint for the EPED model. It will be shown that this can be accomplished even with a low velocity resolution. The implementation of GFS in EPED is described in a separate paper \cite{anastopoulos:2025}.

% The second goal of this study is to determine if GFS is able to be used for a quasi-linear transport model of the pedestal region. The NSTX plasma profiles in the database include separate electron and ion temperature and density plus a carbon impurity species. The database of CGYRO runs only has linear growthrates and frequencies at low poloidal wavenumbers since it was focused on KBM. A full test of GFS for transport would require non-linear gyrokinetic flux tube simulations including electron scales. This is not yet available. The conclusion of this study is that at the velocity space resolution needed to compute the full transport system (particle, energy, momentum) using GFS, with Lorentz pitch angle scattering collisions, is accurate enough but it suffers from a small 2.5\% number of cases that have very high frequencies compared to CGYRO. These outliers are investigated further in this paper and found to be tearing-parity MTM like modes. 

The second goal of this study is to determine if GFS is able to be used for a quasi-linear transport model of the pedestal region. The NSTX plasma profiles in the database include separate electron and ion temperature and density plus a carbon impurity species. The database of CGYRO runs only has linear growth rates and frequencies at low poloidal wavenumbers since it was focused on KBM. A full test of GFS for transport would require non-linear gyrokinetic flux tube simulations including electron scales. This is not yet available. The conclusion of this study is that at the velocity space resolution needed to compute the full transport system (particle, energy, momentum) using GFS, with Lorentz pitch angle scattering collisions, is accurate enough for quasi-linear transport applications. The CGYRO analysis identified KBM, MTM and trapped electron modes (TEM) in the pedestal region as the most unstable mode for different plasma parameters. The GFS linear eigenmode code finds the most unstable mode and subdominant modes that can be a better match to the most unstable mode found with CGYRO. Hence we select the closest eigenvalue mode branch to CGYRO instead of the most unstable mode from GFS to compute the eigenvalue error and keep track of the percent of cases where the GFS branch selected was not the most unstable.

 % We introduce a novel Bayesian optimization approach to determine optimal parallel Hermite-basis resolution at different velocity resolutions, balancing computational efficiency with physical accuracy for spherical tokamak pedestal conditions. Our analysis focuses on ELMy NSTX discharge 139047 \cite{Diallo2013}, finding both kinetic-ballooning and trapped electron modes are unstable across the H-mode pedestal region parameter database.

We introduce a novel Bayesian optimization approach to determine optimal parallel Hermite-basis resolution at different velocity resolutions, balancing computational efficiency of running GFS at lower velocity resolution than CGYRO with physical accuracy. Our analysis focuses on ELMy NSTX discharge 139047 \cite{Diallo2013}, finding that kinetic-ballooning, trapped electron, and micro-tearing modes are unstable across the H-mode pedestal region parameter database. The optimization methodology employs a closest eigenvalue selection strategy that minimizes the combined growth rate and frequency error relative to CGYRO results, significantly improving mode identification accuracy.

% Key findings demonstrate that high collisionality in NSTX pedestals enables reduced parallel velocity resolution in GFS without sacrificing accuracy in linear stability analysis, leading to significant computational savings. The optimized GFS configurations substantially outperform standard TGLF \cite{Staebler:2021} models in both growth rate and frequency predictions. However, we also identify specific parameter regimes where mode identification challenges persist, providing guidance for future model development.
% This work represents a significant advancement in extending reduced transport models to low aspect-ratio configurations while maintaining computational efficiency through systematically optimized resolution parameters. 

Key findings demonstrate that the optimized GFS resolution configuration achieves exceptional accuracy with RMS deviation from CGYRO of 15.91\% (growth rate) and 21.11\% (frequency) with 10\% mode identification mismatch. The GFS linear eigensolver substantially outperforms the TGLF linear eigensolver. However, we identify specific parameter regimes where challenges persist for GFS: low magnetic shear conditions ($\hat{s} < 5$) and near the separatrix ($r/a > 0.97$), where extreme gradients and electromagnetic effects degrade eigenmode accuracy.

The paper is structured as follows: Section \ref{sec:data} details the selected NSTX discharge and data analysis methods. Section \ref{sec:gfs} describes the GFS transport code and its advantages over other models. Section \ref{sec:numeric} presents linear stability analysis results and Bayesian optimization for eigenmode resolution selection.  Section \ref{sec:tglf} compares optimized GFS performance against TGLF. Section \ref{sec:analyerror} provides parametric analysis of eigenvalue RMS deviations to identify where GFS performs well versus where challenges remain.  Section \ref{sec:conclusion} concludes with key findings and discussion.

\section{NSTX data details}
\label{sec:data}
Previous gyrokinetic stability analysis with the CGYRO code \cite{parisi2024kinetic} of the ELMy H-mode NSTX discharge 139047 at 600ms \cite{Diallo2013}, exhibits various instabilities across different wave number ranges \cite{parisi2024kinetic}. However, KBMs was identified as the primary mechanism limiting pedestal structure. Local gyrokinetic analysis at three radial pedestal locations showed that experimental profiles operate near KBM thresholds but below ideal-ballooning stability boundaries. This was shown to explain the distinctive pedestal width scaling observed in NSTX compared to conventional aspect ratio tokamaks.

The high-$\beta$ operation characteristic of low aspect ratio configurations introduces electromagnetic effects that drive both MTMs and KBMs. KBMs specifically affect the mid-pedestal region with equal impact on particle and thermal transport $(D_s/\chi_s \sim 1)$, distinguishing them from other instabilities like MTM that primarily produce electron energy transport. 

While the NSTX plasma exhibits multiple instability modes (ITG, TEM, MTM, KBM, and ETG), our GFS verification focuses primarily on KBMs with limited investigation of TEMs and MTMs. Since this work compares GFS and CGYRO, we require the CGYRO results to serve as converged and accurate ground truth. We filter the dataset to retain only cases where the relative difference between the final two CGYRO times series outputs is less than 1\%. The resulting NSTX database comprises 864 cases: 706 KBM cases, 89 MTM cases, and 69 TEM cases, all classified based on CGYRO simulations. Even though only a single discharge is used for the base case, variations in the equilibrium pressure profile were made, and new MHD equilibria computed, to generate a larger database that maps out the diagram of magnetic shear vs Shafranov shift used to find the KBM stability boundary. 

The parameter ranges for this database are presented in Table~\ref{tab:nstx}. For each local parameter the mean value, standard deviation, minimum and maximum values for the whole database are given. These pedestal local plasma conditions are extreme compared to typical core plasma parameters. The inverse gradient lengths ($a/L_{n*},a/L_{T*}$) are ten times the typical core values as are electron-electon collisions ($\nu_{ee}$) and magnetic shear ($\hat{s}$). 

\begin{table}[h!]
\centering
\caption{Database for the NSTX parameters}
\label{tab:nstx}
\begin{tabular}{l S[table-format=2.4] S[table-format=2.4] l}
\toprule
\textbf{Parameter} & {\textbf{Mean}} & {\textbf{Std}} & \textbf{[Min, Max]} \\
\midrule
$q$            & 8.1134  & 0.9856  & [5.3864, 10.6100]   \\
$\hat{s}$            & 8.2762  & 2.5191  & [-0.7035, 21.7390]  \\
$r/a$         & 0.9654  & 0.0153  & [0.9194, 0.9916]   \\
$\nu_{ee} a/cs$       & 7.3118  & 4.4780  & [1.5049, 19.2890]  \\
$Z_{\rm{eff}}$       & 1.8838  & 0.2548  & [1.5505, 2.3359]   \\
$a/L_{T_D}$    & 8.1223  & 8.6647  & [-37.6280, 35.9630] \\
$a/L_{T_C}$    & 8.1223  & 8.6647  & [-37.6280, 35.9630] \\
$a/L_{T_e}$    & 16.9654 & 7.0072  & [5.0209, 54.7530]  \\
$a/L_{n_D}$   & 2.8161  & 3.1670  & [-1.6732, 13.9140]  \\
$a/L_{n_C}$    & 37.0499 & 18.9688 & [12.5360, 126.2600] \\
$a/L_{n_e}$    & 8.3435  & 4.2911  & [2.5493, 26.7960]  \\
$k_y$           & 0.1530  & 0.0380  & [0.06, 0.18]   \\
\bottomrule
\end{tabular}
\end{table}

For this discharge the global parameters are $B_t= 0,45T , R/a= 1.27, \kappa = 2.3, \delta_{upper} = 0.4 $. The line average density is $6 10^{13}/cm^3$. The discharge is heated with neutral beam injection of 6MW. Input files for CGYRO were generated from plasma profiles fit to measured data for electron and ion temperatures, electron data, carbon density at 12 flux surface locations from the pedestal top to the separatrix $r/a\approx 0.93-0.99$. An MHD equilibrium was computed with the EFIT-AI code \cite{Lao_2022} including the bootstrap current \cite{Sauter_1999} from the kinetic pressure gradient. The Miller extended harmonic geometry \cite{arbon_2021} that fits up/down asymmetric flux surfaces was used for the CGYRO and GFS calculations.

\section{Linear gyrokinetic stability models: comparative analysis of gyrokinetic and gyrofluid approaches}\label{sec:gfs}
Numerical analysis of plasma stability requires sophisticated modeling approaches that can accurately capture the complex physics of tokamak plasmas. The gyrokinetic system of equations \cite{catto1978linearized,Frieman1982} is solved by codes such as CGYRO \cite{candy2016high}. The CGYRO code has been extensively bench marked against other high fidelity gyrokinetic solvers and serves as a benchmark for reduced models. The CGYRO initial value linear stability runs were executed at high resolution: with 96 poloidal gridpoints per period, 8 periods, fully electromagnetic, 8 energy quadrature points, 15 pitch angle quadrature points, 1 toroidal mode number with kinetic electrons and 2 ion species (CGYRO inputs: N\_THETA=96, N\_RADIAL=8, N\_FIELD=3, N\_ENERGY=8, E\_MAX=8, N\_XI=15, N\_TOROIDAL=1, N\_SPECIES=3, COLLISION\_MODEL=4). The Sugama multi-species collision model \cite{Sugama:2009} was used for CGYRO. The TGLF and GFS codes only have pitch angle scattering and some differences are expected in the linear eigenvalues due to the different collision models \cite{Belli_2017} at the collision frequencies of the NSTX pedestal database.

The Trapped Gyro-Landau Fluid (TGLF) \cite{Staebler:2005} linear eigensolver solver and quasi-linear transport model has been widely applied to core plasma simulations. TGLF implements bounce-averaged trapped particles with three moments and circulating particles with twelve moments using a generalized Beer-Hammett closure scheme \cite{beer1996toroidal}. As demonstrated in \cite{staebler2007theory}, TGLF has been successfully validated against conventional tokamak data from DIII-D, JET, and ASDEX-Upgrade \cite{angioni2022confinement}. However, previous work \cite{kinsey2025analysis} identified significant limitations in TGLF when applied to low-aspect ratio and high-beta configurations typical of spherical tokamaks. This study revealed that TGLF lacks sufficient perpendicular energy resolution to accurately capture the effects of parallel magnetic fluctuations ($\tilde{B}_{||}$) that are critical for kinetic ballooning mode (KBM) stability in low-aspect ratio devices. The velocity resolution of the 15 moments equations per species in TGLF cannot be changed due to the complex machine learning fit closure scheme. 

To address these and other limitations, the Gyro-Fluid System (GFS) solver \cite{staebler2023flexible} has been developed with several key improvements over previous gyrofluid models. Unlike TGLF, GFS does not use a bounce-averaging approximation, but has the mirror force in the gyrokinetic equations. This enables GFS to capture the trapped electron mode even at low aspect ratio and high trapped fraction. A flexible velocity space spectral method employs moments of the gyroknetic equation with a variable number ($nu$) of Hermite polynomials in the parallel velocity and Laguerre polynomials in the perpendicular energy with a variable number ($ne$). A two parameter real closure of the Hermite polynomial moments produces sufficient accuracy even at much lower velocity space resolution compared to standard pseudo-spectral methods used in gyrokinetic codes. Both GFS and TGLF use a Hermite polynomial representation of the straight magnetic field line spatial coordinate ($z$) with a variable width ($wz$) of the Gaussian measure and a variable number of polynomials ($nz$). Using a database of core NSTX and NSTX-U CGYRO linear stability runs \cite{kinsey2025analysis} demonstrated that GFS captures the strong destabilizing effect of $\tilde{B}_{||}$ on KBMs predicted by CGYRO, an effect that TGLF struggles to reproduce even with frequency filtering adjustments.

Systematic benchmarking of GFS with CGYRO for parameter scans around an NSTX-U core standard case showed remarkable agreement across electromagnetic parameter scans \cite{kinsey2025analysis}. For electrostatic cases, GFS achieved an average RMS error of 15.7\% compared to CGYRO, substantially outperforming TGLF-SAT1 (44.4\%) and TGLF-SAT2 (32.2\%) versions with different collision models. This improvement was even more pronounced for fully electromagnetic cases including parallel magnetic fluctuations, where GFS maintained an error of 29.2\% while TGLF-SAT1 and TGLF-SAT2 exhibited errors of 97.4\% and 52.8\%, respectively. These verification results establish GFS as a significant advancement for modeling spherical tokamak core plasmas at higher $\beta$ where electromagnetic effects are critical.

While previous work validated GFS for core ST scenarios \cite{kinsey2025analysis}, our current work extends GFS to pedestal applications, which present distinct challenges due to extremely steep gradients and strong electromagnetic effects. Because numerical resolution requirements for gyrokinetic simulations are typically much more demanding in the pedestal than the core \cite{Parisi2020b}, successful validation of GFS is also expected to require different resolution settings than used in the core. The GFS settings for the NSTX pedestal database will be determined by Bayesian optimization in this paper.

\section{Linear stability analysis with the GFS model}\label{sec:numeric}
 
In this study, we present a systematic benchmark of GFS across multiple resolution settings to establish optimal parameters for NSTX pedestal conditions. We begin by testing the established core-optimized resolution \cite{staebler2023flexible} of three perpendicular energy moments ($ne = 3$), seven parallel velocity moments ($nu = 7$), a Gaussian width of $wz=1.74$ and  $nz=12$ parallel z-moments. These core-optimized resolution settings were also shown to be sufficient for drift-waves and KBM in NSTX-U.  Subsequently, we employ Bayesian optimization to determine optimal $(wz, nz)$ parameters for different velocity resolution settings. For each configuration, we compare  the eigenvalues from GFS to the initial value linear stability results from CGYRO.

\subsection{Evaluation of core-optimized configuration for pedestal conditions}
\begin{figure}[h!]
    \centering    \includegraphics[width=0.99\textwidth]{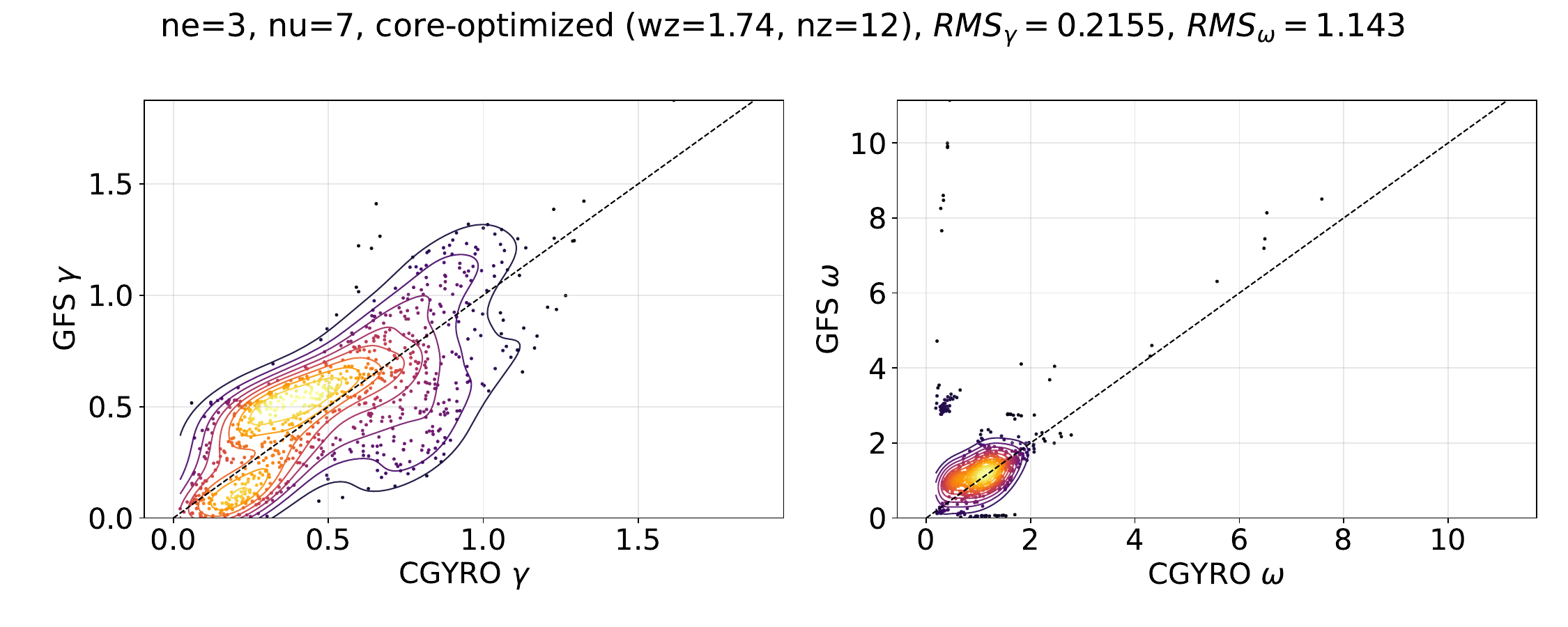}
    \caption{Comparison of GFS and CGYRO growh rates (left) and frequencies (right) using the most unstable GFS mode with core-resolution settings ($ne=3$, $nu=7$, $wz=1.74$, $nz=12$), showing RMS error of growth rate $\gamma \sim 21.55\%$ (left) and RMS error of frequency $\sim 114.30\%$ (right).}
    \label{fig:eigs_optim_ne3_nu7_default}
\end{figure}

The first step is to examine GFS performance using the established core-resolution configuration parameters: $ne=3$, $nu=7$, $wz=1.74$, and $nz=12$. Fig.~\ref{fig:eigs_optim_ne3_nu7_default} presents a comparison between GFS with the core tokamak resolution and CGYRO for NSTX pedestal conditions for the whole CGYRO database. The analysis reveals a 21.55\% Root Mean Square (RMS) error in growth rate and a 114.30\% RMS error in frequency. Here, the RMS error ${\rm Err}_{\rm rms}(x)$ of a $N$-size database is defined as ${\rm Err}_{\rm rms}(x) = \sqrt{\frac{1}{N}\sum_{n}^N|x_{\rm gfs}^{(n)} - x_{\rm cgyro}^{(n)}|^2}$ for quantity of interest $x$.

Fig.~\ref{fig:eigs_optim_ne3_nu7_default} illustrates that while the core-resolution demonstrates promising agreement between GFS and CGYRO, several limitations remain evident for pedestal applications. The growth rate ($\gamma$) predictions show reasonable concordance with RMS error around 20\%, particularly in the lower range ($\gamma < 0.75$). However, the frequency ($\omega$) predictions exhibit high frequency outliers in the low CGYRO $\omega$ region. To fully understand the source of these discrepancies we examine the high frequency outliers.

\subsubsection{Analysis of high frequency outliers and mode identification}\label{sec:outlier}

This section investigates the 9 (1.0\%) frequency outliers in GFS with core-resolution all of which were classified as KBM modes in the CGYRO simulations.

\begin{figure}[h!]
    \centering    
    \includegraphics[height=0.225\textwidth]{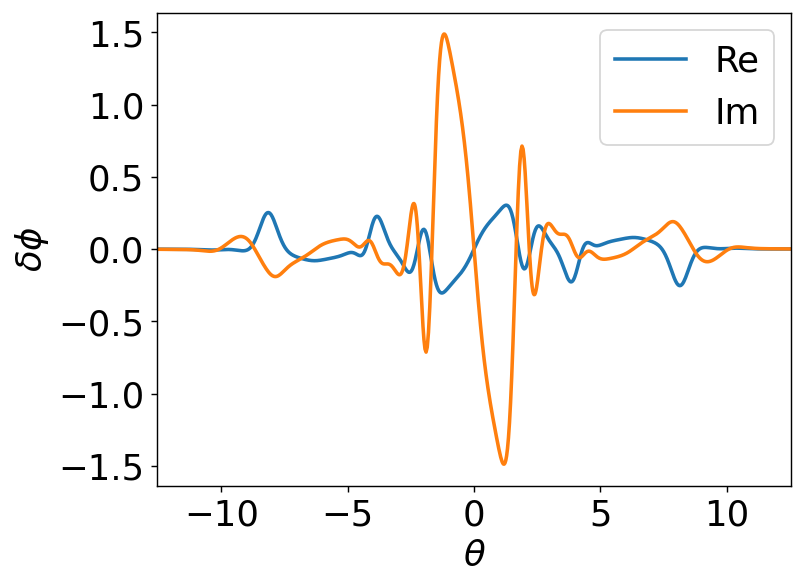} \includegraphics[height=0.225\textwidth]{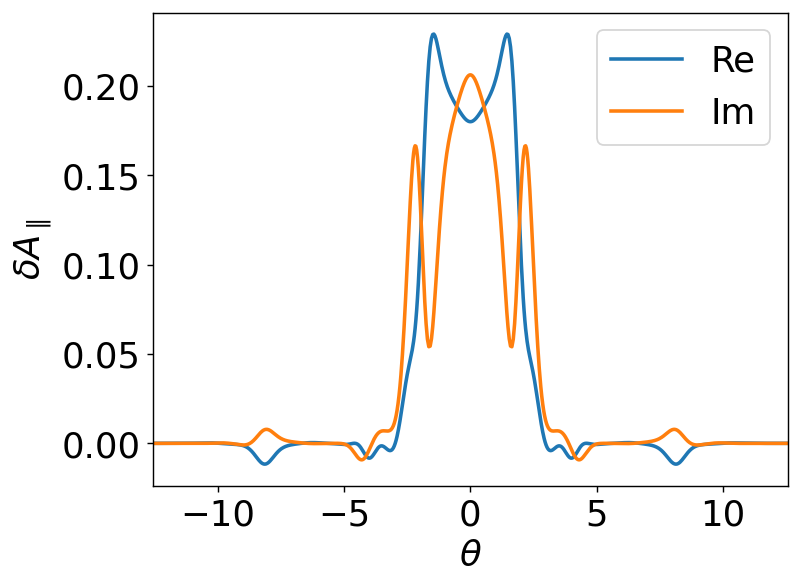} \includegraphics[height=0.225\textwidth]{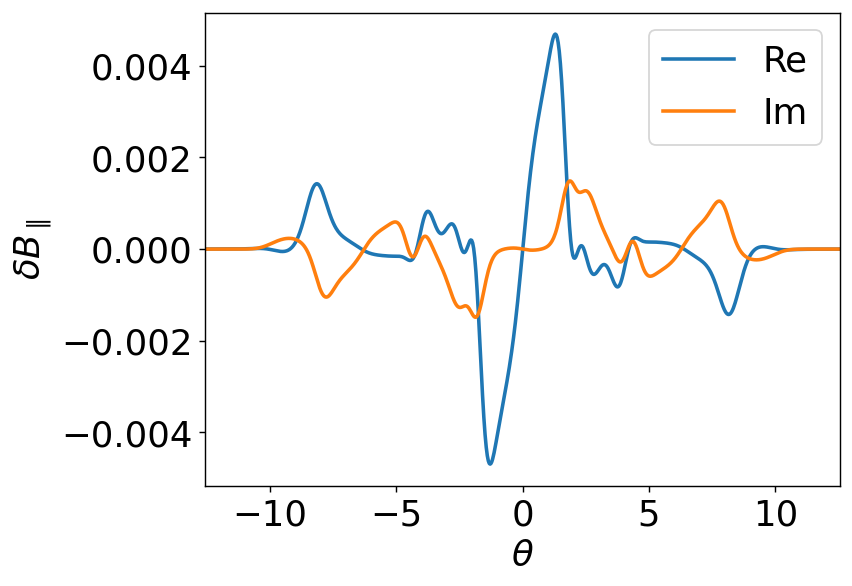}
    \vspace{-0.5cm}
    \caption{Eigenmode structure of high frequency mode. The odd parity in [$\delta\phi$, $\delta B_\parallel$] and even parity in $\delta A_\parallel$ indicates that this mode is not a KBM but an electrostatic odd-parity mode.}
    \label{mtm}
\end{figure}

Although the frequencies are higher (4-10) for GFS than CGYRO (< 1) for these outliers, recall that the gradients are extremely high and these frequencies are not larger than the diamagnetic drift frequencies in the gyrokinetic equation. It is also not unexpected that, at the high collision frequencies of this dataset, the full multi-species collision operator of CGYRO \cite{Belli_2017} that was used for these calculations, would give different eigenvalues than the simple Lorentz pitch angle scattering model in GFS. GFS includes like-species Lorentz collisions for ions and electrons plus the Zeff contribution to electrons from the ions. The outlier GFS modes are found to be tearing-parity modes rather than KBMs. An example of such an eigenmode structure can be seen in Fig.~\ref{mtm}, where a up/down symmetric version of the equilibrium geometry is used to clearly identify the mode parity. The electric potential and parallel magnetic field [$\delta\phi$, $\delta B_\parallel$] are odd parity and the parallel vector potential $\delta A_\parallel$ is even parity. This indicates that this mode is not a KBM but rather a tearing-parity mode like the MTM. These high frequency modes are similar to the ones often found with TGLF close to the KBM threshold in the plasma core. They are thought to be poorly spatially resolved micro-tearing modes \cite{Najlaoui_2025}. These outlier modes were not observed with GFS for core NSTX/NSTX-U plasma conditions.  

\begin{figure}[h!]
    \centering    
    \includegraphics[height=0.225\textwidth]{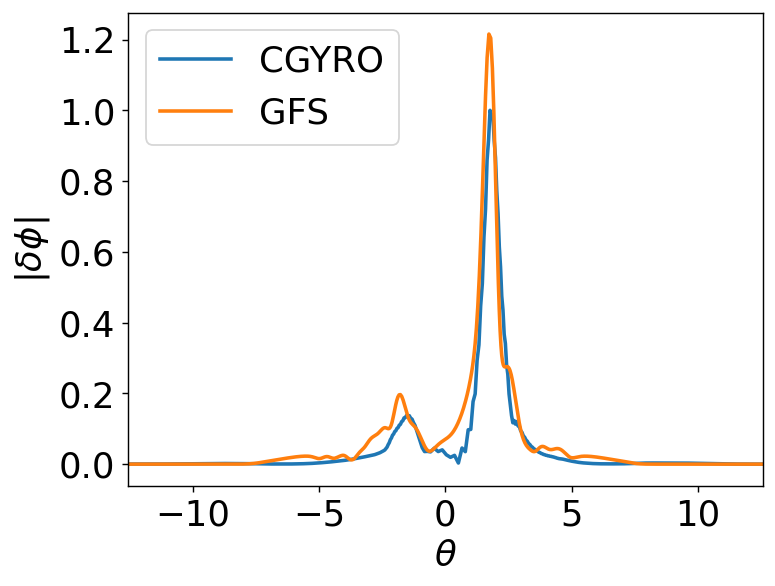} \includegraphics[height=0.225\textwidth]{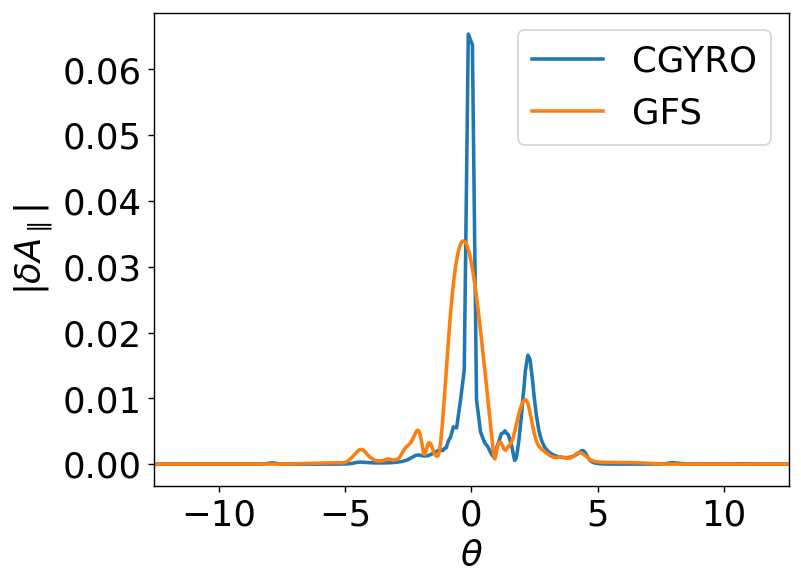} \includegraphics[height=0.225\textwidth]{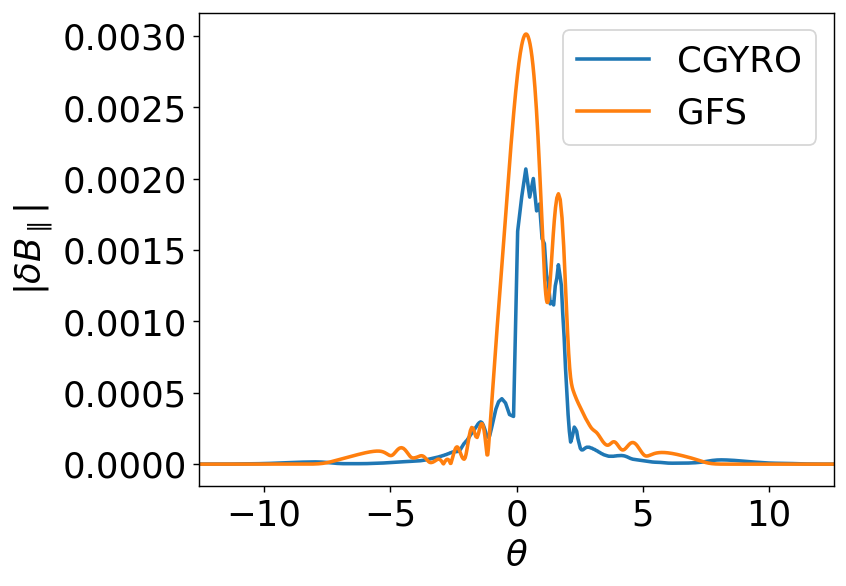}
    \vspace{-0.5cm}
    \caption{Comparison of the CGYRO and GFS eigenmodes $\delta\phi$, $\delta A_\parallel$ and $\delta B_\parallel$ for the subdominant KBM case.}
    \label{gfs_cgyro_mode}
\end{figure}

\begin{figure}[h!]
    \centering    
    \includegraphics[width=0.39\textwidth]{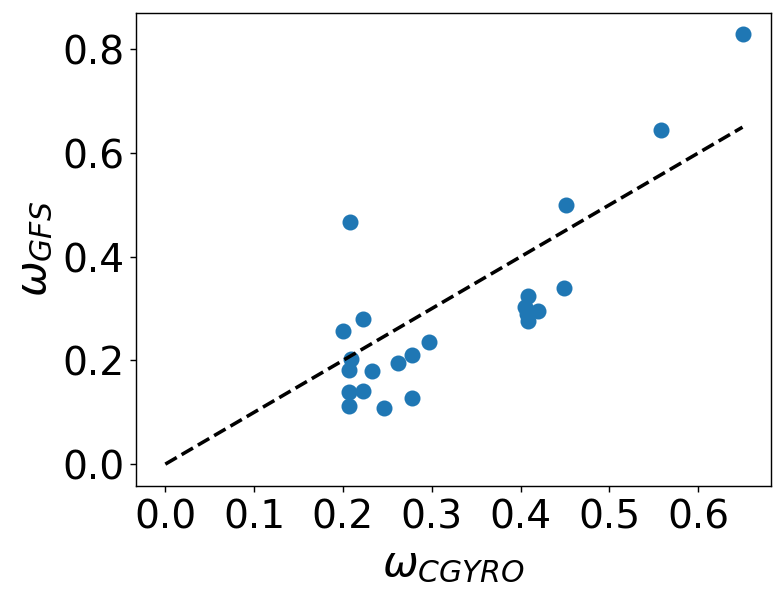} \ \ \ \ \includegraphics[width=0.4\textwidth]{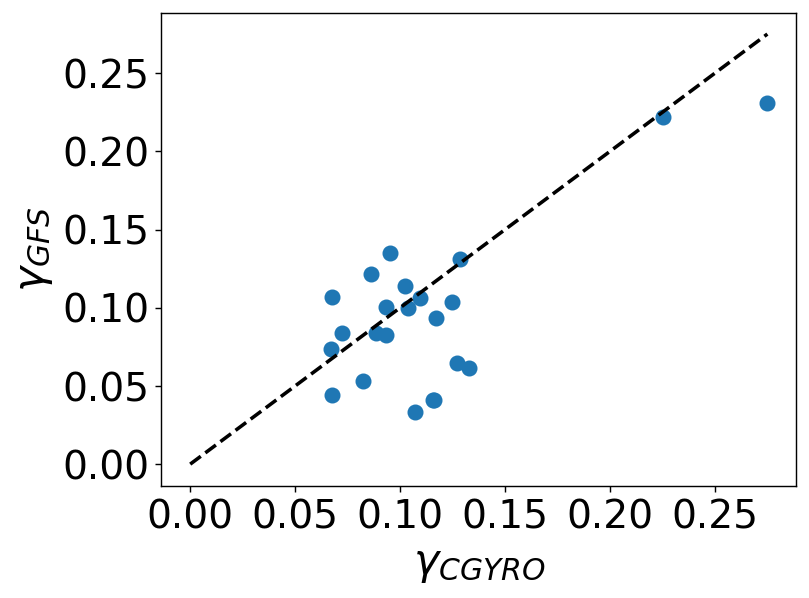} 
    \vspace{-0.5cm}
    \caption{Comparison of CGYRO and GFS frequency (left) and growth rate (right) for cases where GFS identifies KBMs as the subdominant instability.}
    \label{omega_comp}
\end{figure}

Since GFS is an eigenvalue solver, it is possible to retrieve subdominant eigenmodes that better correspond to the CGYRO results. To verify that these subdominant modes are indeed the relevant modes for comparison, we compare the three electromagnetic potentials for GFS to CGYRO, which results in good agreement as shown in Fig.~\ref{gfs_cgyro_mode}. When considering only these subdominant KBM cases, the relative errors for the growth rate $\gamma$ and the frequency $\omega$ between CGYRO and GFS (at the core resolution) are $27\%$ and $31\%$ respectively, with Fig.~\ref{omega_comp} showing the comparison of the eigenvalues.

Excluding the high frequency simulation cases that are not classified as KBM cases by GFS, the RMS error analysis shows substantially improved results, with an RMS error of growth rate at approximately 14.69\% and an RMS error of frequency at approximately 26.58\%. These results demonstrate competitive accuracy in growth rate prediction while achieving significantly better performance in frequency estimation when appropriate mode identification is applied.

\subsubsection{GFS with Selected Closest Branch}

Based on the analysis of high-frequency outliers, the most unstable mode from GFS may not correspond to the CGYRO result. To address this issue, we calculate GFS with three unstable modes ($n_{\rm mode}=3$) and select the optimal/closest (minimal eigenvalue RMS error) one to the CGYRO for comparison to CGYRO
\begin{equation}
    \{\gamma_{\rm gfs}, \omega_{\rm gfs}\} = \min_{j}\left\{|\gamma_{\rm gfs}^j-\gamma_{\rm cgyro}|^2+{|\omega_{\rm gfs}^j-\omega_{\rm cgyro}|^2}\right\}, \quad j=1, \ldots, n_{\rm mode}.
\end{equation}
To quantify mode identification mismatch, we introduce the variable $\sigma_{\rm branch} = \frac{N_{\rm mismatch}}{N_{\rm total}}$, where $N_{\rm mismatch}$ is the number of cases where the most unstable GFS mode was not the closest eigenvalue to CGYRO, to represent the fraction of cases where the optimal GFS eigenvalue is not the most unstable one.

\begin{figure}[h!]
    \centering    
    \includegraphics[width=0.99\textwidth]{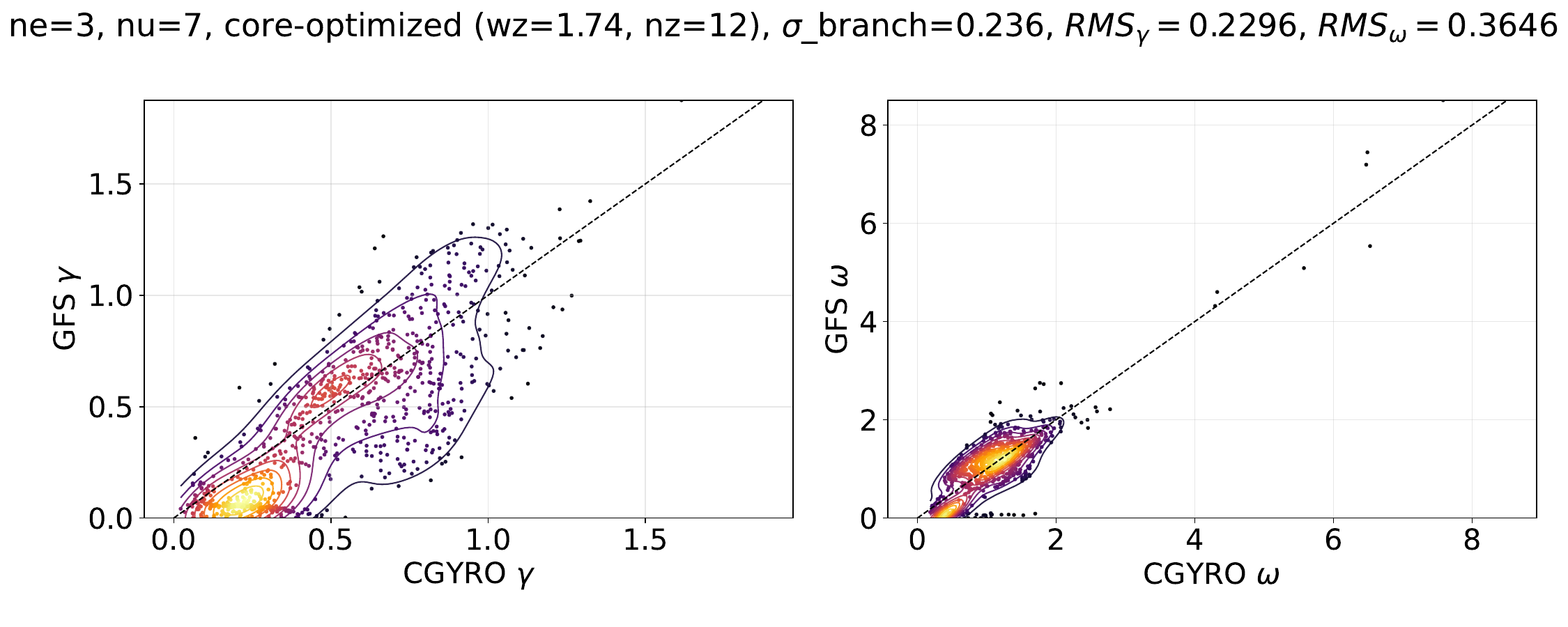}
    \caption{Eigenvalue comparison between CGYRO and GFS with optimal mode branch with core-optimized resolution settings ($ne=3$, $nu=7$, $wz=1.74$, $nz=12$). The growth rate RMS error is 22.96\% and frequency RMS error is 36.46\%. Mode identification mismatch ratio $\sigma_{\rm branch} = 23.6\%$.}
    \label{fig:eigs_optim_ne3_nu7_default_closest}
\end{figure}

Fig.~\ref{fig:eigs_optim_ne3_nu7_default_closest} shows the eigenvalue comparison between CGYRO and GFS using the closest eigenvalue mode branch. The results demonstrate that high-frequency outliers are eliminated when choosing the optimal mode branch instead of the most unstable mode. The frequency RMS error decreases dramatically to 36.46\%, a significant improvement over the 114.30\% error obtained with the most unstable mode. However, the mode identification mismatch rate of 23.6\% ($\sigma_{\rm branch} = 0.236$) remains problematic for practical applications.

This analysis clearly demonstrates that the core-resolution performs inadequately in the pedestal region, exhibiting both relatively high eigenvalue errors and  mode identification mismatch. Therefore, the next step involves using Bayesian optimization to investigate optimal spatial grid parameters ($wz, nz$) for different velocity resolutions.

\subsection{Bayesian optimization methodology}

Bayesian optimization is employed to systematically determine optimal resolution settings for the GFS model applied to NSTX pedestal conditions, specifically targeting two key hyperparameters: the number of Hermite basis functions along the magnetic field ($nz$) and the Gaussian z-width ($wz$).
In our optimization framework, for fixed $ne$ and $nu$, we search the parameter space $(nz,wz) \in ([10,44], [0.8,4.0])$. The optimization process utilizes Gaussian process regression \cite{schulz2018tutorial} to construct a surrogate model of the objective function landscape. This surrogate model continuously updates its predictions of the objective function as new evaluations are performed, incorporating uncertainty estimates to guide the search process.

The cost function $\mathcal{L}$ considers both growth rate and frequency errors between GFS using the closest eigenvalue mode branch and CGYRO, which is assumed as the ground-truth solution:
\begin{equation}
   \mathcal{L} 
   = -\sqrt{\frac{1}{I_{\rm NSTX}} \sum_{i=1}^{I_{\rm NSTX}}\min_{j}\left\{ |\gamma_{\rm gfs}^{i,j} - \gamma_{\rm cgyro}^{i}|^2 + |\omega_{\rm gfs}^{i,j} - \omega_{\rm cgyro}^{i}|^2  \right\}},\quad j=1, \ldots, n_{\rm mode},
\end{equation}
where $n_{\rm mode}=3$ for GFS simulations, and $I_{\rm NSTX}$ represents the number of NSTX simulation cases used for benchmarking. To balance the optimization requirements for each mode type, we subset the data and equally select 40 KBMs (out of 706), 30 TEMs (out of 69), and 40 MTMs (out of 89). This balanced sampling prevents the optimization from being biased toward KBMs, which dominate the full dataset ($\approx 82\%$). The objective is to maximize the cost function $\mathcal{L}$, which effectively minimizes the RMS error between CGYRO and GFS using the closest/optimal eigenvalue mode. We monitor the mode mismatch rate $\sigma_{\rm branch}$ during optimization but do not include it in the cost function.

The optimization implementation uses an open-source Bayesian optimization library \cite{code_bay}, treating the GFS simulation as a black-box function. This function accepts input parameters ($nz$, $wz$) within their specified ranges and produces a scalar output (the cost function). The search process runs for 40 iterations, initialized with 2 random points to establish a baseline surrogate model. The Expected Improvement (EI) acquisition function guides subsequent sampling point selection, balancing exploration of uncertain regions with exploitation of promising areas. There is an even integer constraint on $nz$ (to avoid a zero eigenvalue of $k_z$) that is handled through specialized discrete parameter techniques \cite{garrido2020dealing}, ensuring efficient search characteristics while respecting the discrete nature of this parameter.

This systematic approach enables efficient exploration of the parameter space while accounting for the computational cost of each GFS evaluation, ultimately identifying parameter combinations that optimize the model's performance for NSTX pedestal stability analysis.

\subsubsection{Bayesian optimization for searching optimal $nz$ and $wz$}

\begin{figure}[h!]
    \centering
 \includegraphics[width=0.95\textwidth]{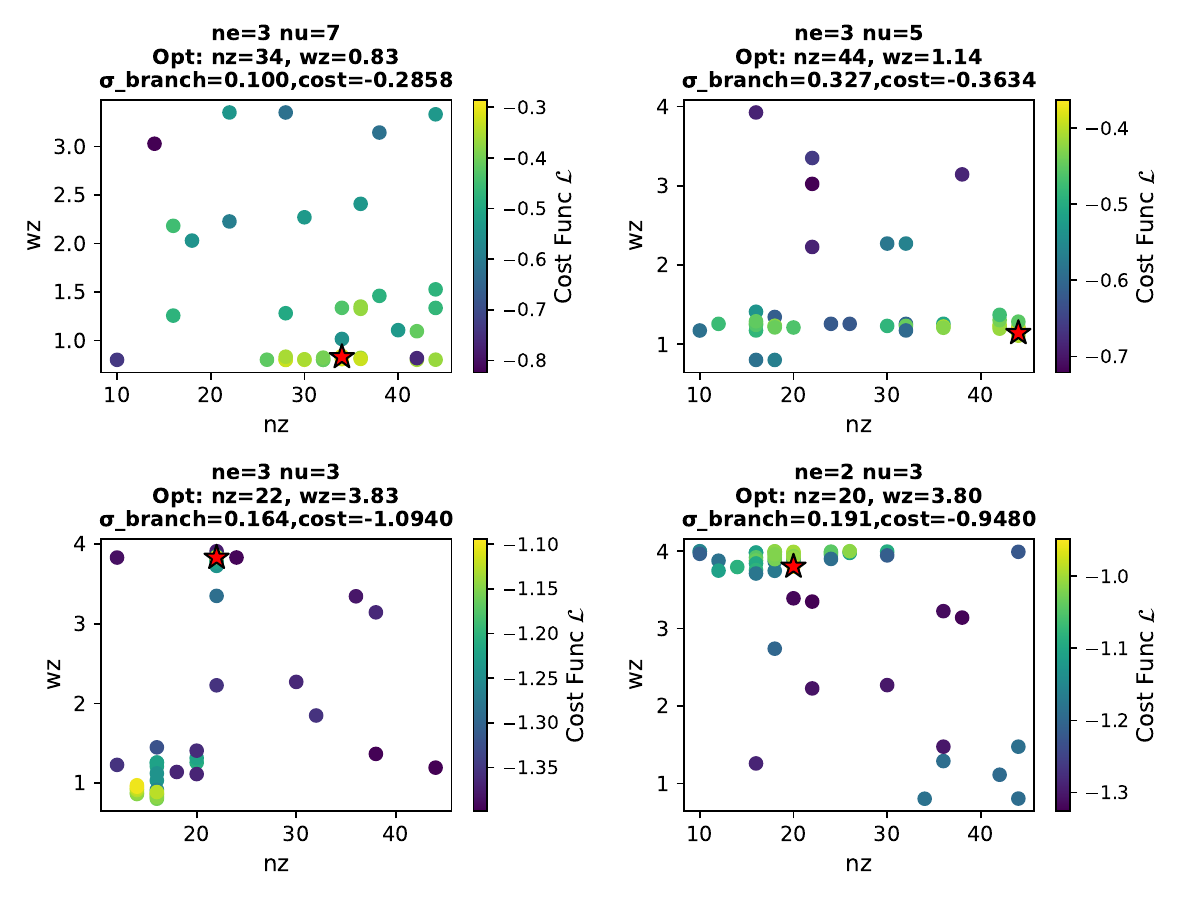}
    \caption{Bayesian optimization results showing scatter plots of $(nz, wz)$ colored by cost function $\mathcal{L}$. Red stars indicate optimal $\mathcal{L}$ values from 42 search points each. (a) $ne=3, nu=7$: optimal $\mathcal{L} = -0.2858$ at $(nz,\,wz) = (34,\,0.83)$. (b) $ne=3, nu=5$: optimal $\mathcal{L} = -0.3634$ at $(nz,\,wz) = (44,\,1.14)$. (c) $ne=3, nu=3$: optimal $\mathcal{L} = -1.094$ at $(nz,\,wz) = (22,\,3.83)$. (d) $ne=2, nu=3$: optimal $\mathcal{L} = -0.948$ at $(nz,\,wz) = (20,\,3.80)$. The configuration $ne=3, nu=7$ with $(nz,\,wz) = (34,\,0.83)$ achieves the best overall performance (lowest eigenvalue error and mismatch fraction).}
    \label{fig:optim_eig}
\end{figure}

Fig~\ref{fig:optim_eig} shows the Bayesian optimization results for four velocity resolution scenarios. The results demonstrate a significant dependence of model performance on parallel velocity moments ($nu$) rather than energy resolution ($ne$). Configurations with higher $nu$ values achieve superior performance: ($ne=3$, $nu=7$) and ($ne=3$, $nu=5$) yield comparable results with cost function values of $\mathcal{L} = -0.2858$ and $\mathcal{L} = -0.3634$, respectively. Performance degrades for lower $nu$, with ($ne=3$, $nu=3$) showing $\mathcal{L} = -1.094$ and ($ne=2$, $nu=3$) achieving $\mathcal{L} = -0.948$. Notably, the optimal case for ($ne=3$, $nu=7$) resolution achieves both excellent eigenvalue accuracy and lowest mode branch identification mismatch  $\sigma_{\rm branch}=0.1$. However, the high optimum $nz=34$ slows the GFS calculation by a factor of 8 compared to $nz=12$. Even for this higher spatial resolution GFS is far faster than CGYRO. The velocity resolution for GFS is lower in both parallel velocity (7/16) and energy (3/8 ) than CGYRO and the number of spatial moments (34) is far smaller than the number of spatial gridpoints (8 x 96 = 768) used in CGYRO. If GFS used the same resolution as CGYRO it would be 18,956 times slower than the optimized predestal-resolution for GFS. 

\begin{figure}[h!]
    \centering    
        \includegraphics[width=0.99\textwidth]{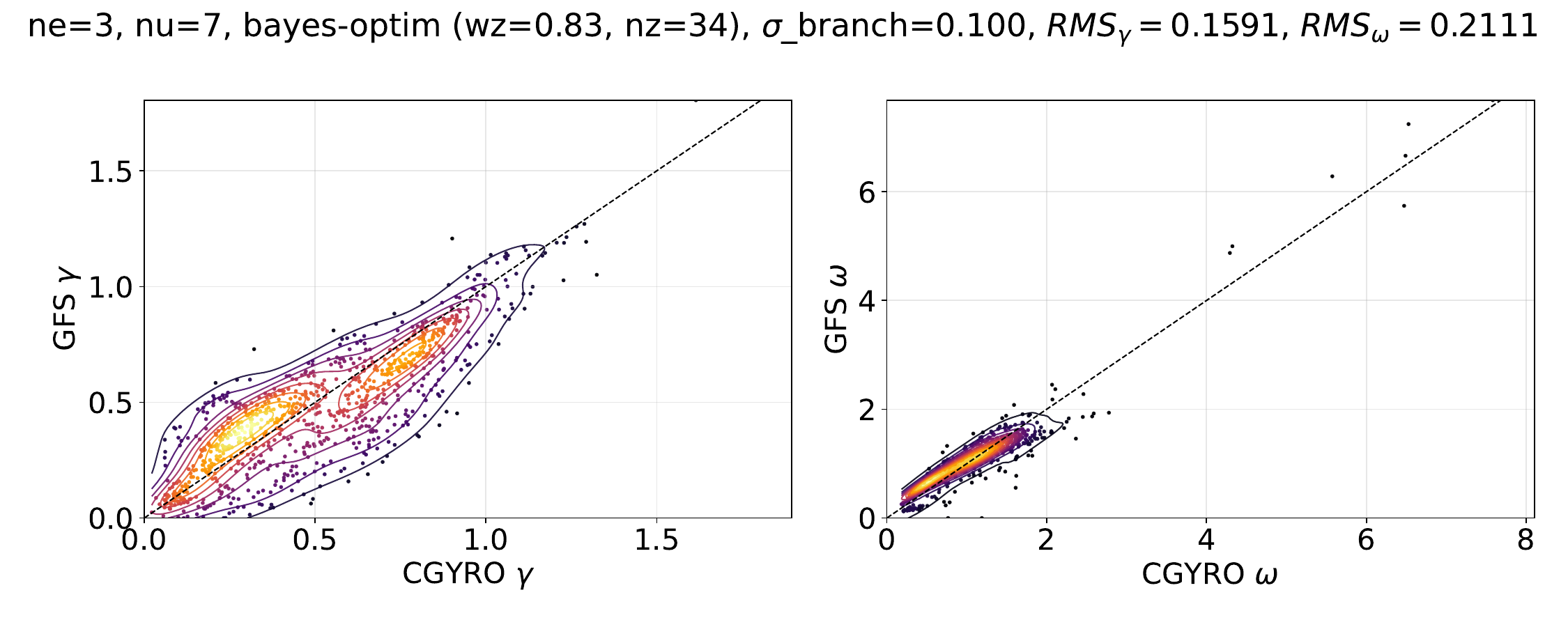}
    \caption{Eigenvalue comparison between CGYRO (x-axis) and GFS (y-axis) for the optimized $(ne=3, nu=7, wz=0.83, nz=34)$ configuration, showing RMS error of growth rate $\gamma \sim 15.91\%$ and RMS error of frequency $\sim 21.11\%$.}
    \label{fig:eigs_optim_ne3_nu7}
\end{figure}

\begin{figure}[h!]
    \centering    
    \includegraphics[width=0.99\textwidth]{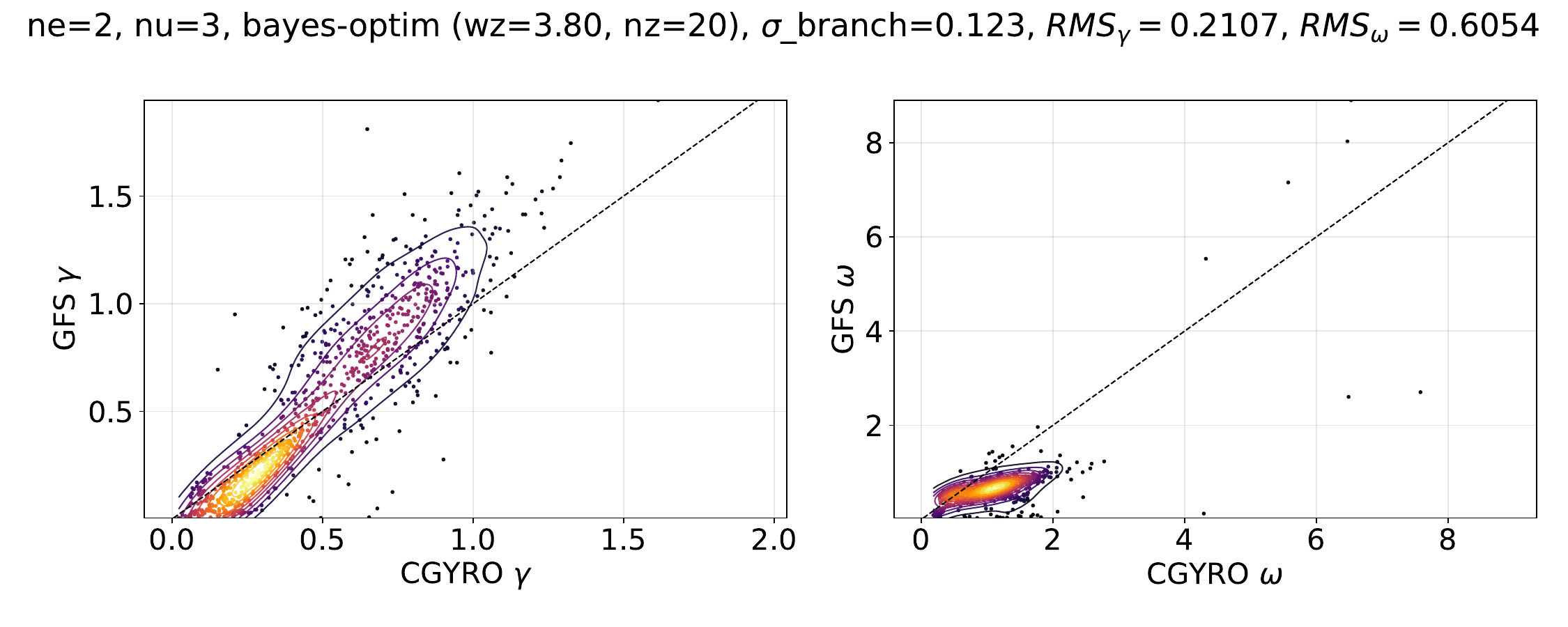}
    \caption{Eigenvalue comparison between CGYRO (x-axis) and GFS (y-axis) for the optimized $(ne=2, nu=3, wz=3.80, nz=20)$ configuration, showing RMS error of growth rate $\gamma \sim 21.07\%$ (left) and RMS error of frequency $\sim 60.54\%$ (right).  }
    \label{fig:eigs_optim_ne2_nu3}
\end{figure}

GFS demonstrates similar performance in eigenvalue calculation between configurations ($ne=3, nu=7$) and ($ne=3, nu=5$), as well as between ($ne=2, nu=3$) and ($ne=3, nu=3$). Based on these results, we focus our detailed eigenvalue comparison on two representative cases: ($ne=3, nu=7$) and ($ne=2, nu=3$) with their respective optimal $(nz, wz)$ values. 
Figs.~\ref{fig:eigs_optim_ne3_nu7} and~\ref{fig:eigs_optim_ne2_nu3} present eigenvalue comparisons for the complete NSTX database using these two configurations. The analysis demonstrates that GFS achieves excellent growth rate predictions across both cases. For frequency predictions, GFS shows good accuracy for the higher velocity resolution ($ne=3, nu=7$) configuration with 21.11\% RMS error. However, the lower velocity resolution case ($ne=2, nu=3$) exhibits systematically lower GFS frequencies compared to CGYRO values, resulting in 60.54\% RMS error. Both configurations maintain acceptable mode branch identification with approximately 10\% mismatch rates.

\subsubsection{Mode-specific performance analysis}

\begin{figure}[h!]
    \centering    
        \includegraphics[width=0.75\textwidth]{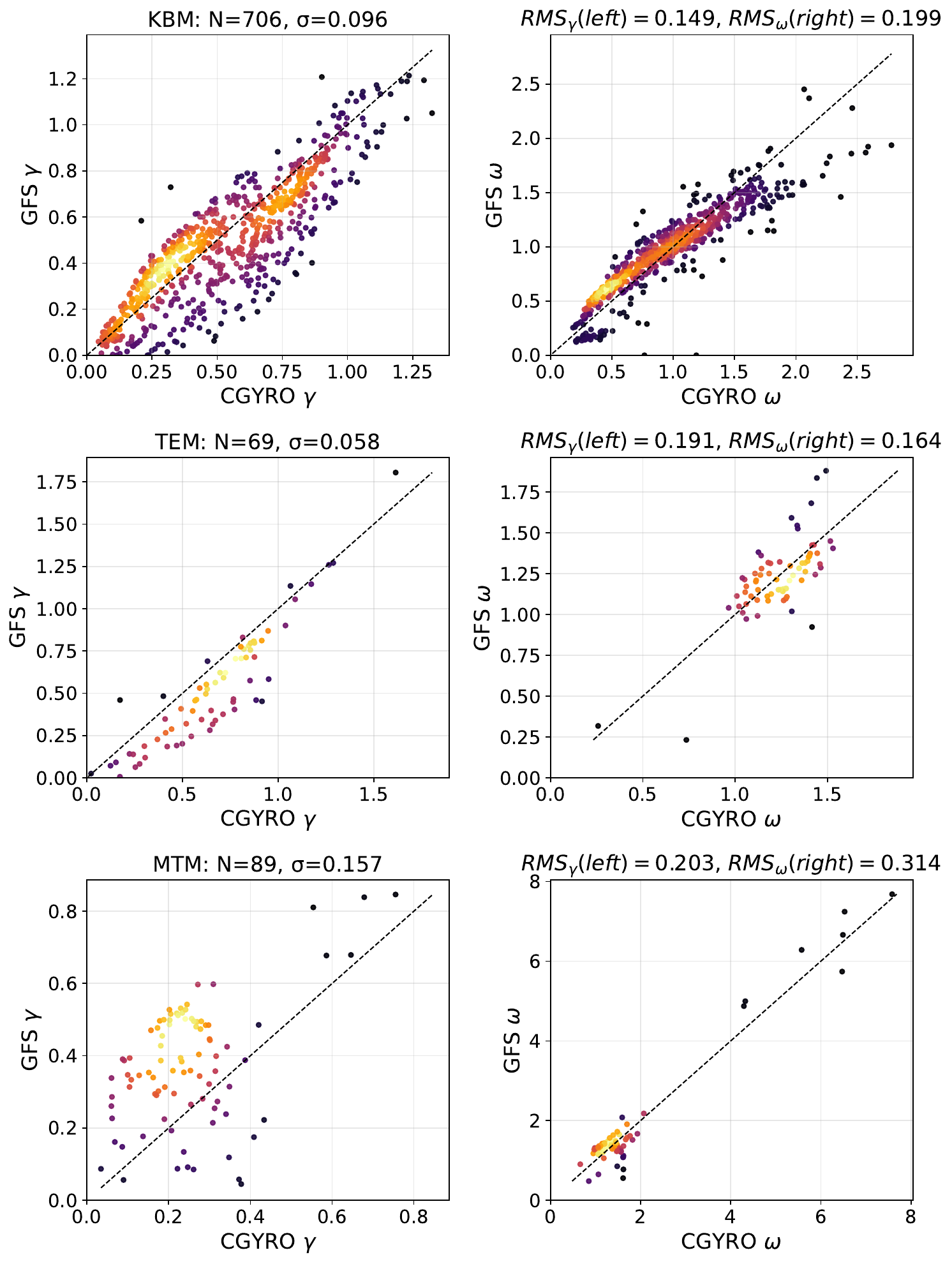}
    \caption{Eigenvalue comparison between CGYRO (x-axis) and GFS (y-axis) for the optimized $(ne=3, nu=7, wz=0.83, nz=34)$ configuration individually for KBM (top), TEM (middle), MTM (bottom). Overall, GFS has excellent performance in KBM and TEM cases and slightly worse in MTM simulation cases.}
    \label{fig:eigs_optim_ne3_nu7_seperate}
\end{figure}

While the (ne=3, nu=7) configuration demonstrates excellent overall performance, detailed mode-specific analysis reveals important variations across different instability types. Fig.~\ref{fig:eigs_optim_ne3_nu7_seperate} presents eigenvalue comparisons for individual mode types using the optimized parameters (wz=0.83, nz=34). GFS achieves excellent performance for both KBM and TEM cases, with RMS errors of 14.9\% (growth rate) and 19.9\% (frequency) for KBMs, and 19.1\% (growth rate) and 16.4\% (frequency) for TEMs. Mode branch identification is also excellent for these modes, with $\sigma_{\rm branch} = 0.096$ for KBMs and $\sigma_{\rm branch} = 0.058$ for TEMs.
However, MTM performance shows some degradation compared to KBM and TEM modes, with RMS errors of 20.3\% (growth rate) and 31.4\% (frequency), along with a higher mode mismatch rate of $\sigma_{\rm branch} = 0.157$. This suggests that MTMs may require different optimal spatial resolution parameters than those found through mixed-mode optimization.

\begin{figure}[h!]
    \centering    
        \includegraphics[width=0.99\textwidth]{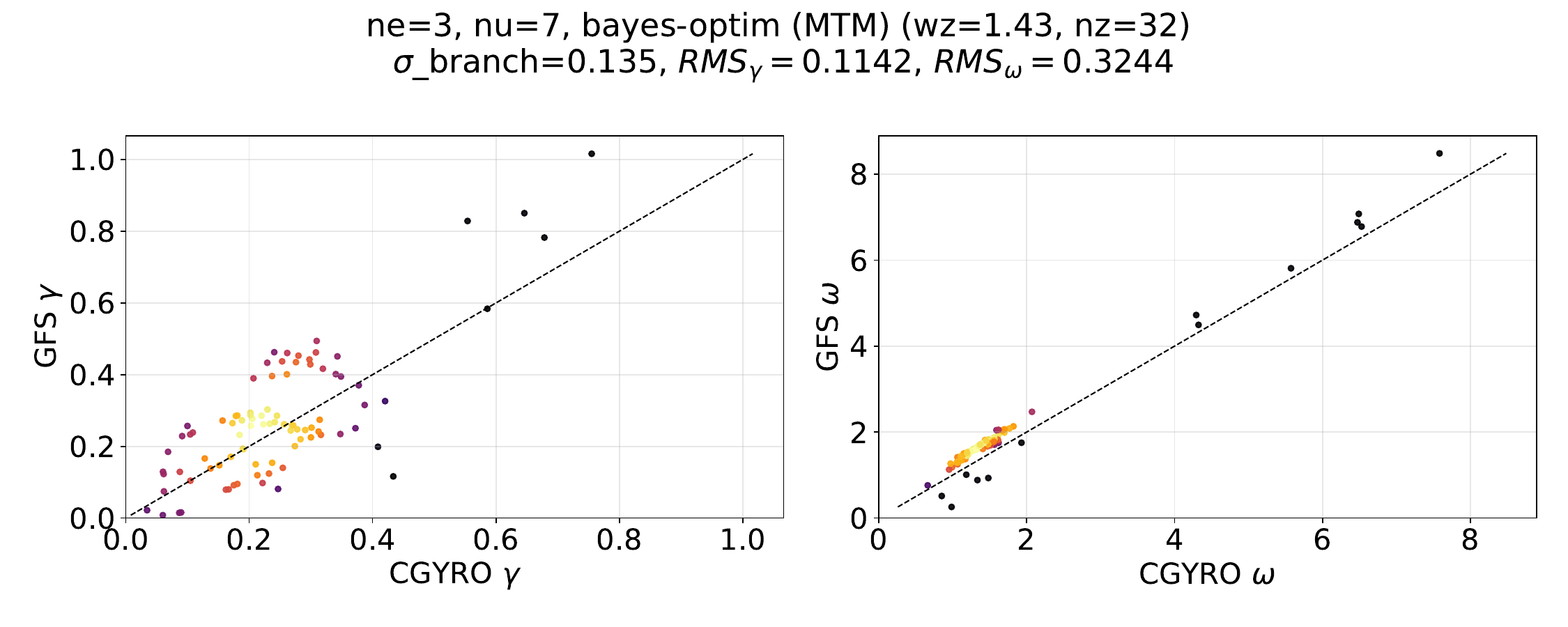}
    \caption{Eigenvalue comparison between CGYRO (x-axis) and GFS (y-axis) for the MTM-optimized $(ne=3, nu=7, wz=1.43, nz=32)$ configuration, showing RMS error of growth rate $\gamma \sim 11.42\%$ and RMS error of frequency $\sim 32.44\%$ with mode identification mismatch ratio $\sigma_{\rm branch} = 13.5\%$.}
    \label{fig:eigs_optim_ne3_nu7_mtm}
\end{figure}

To test this hypothesis, we performed separate Bayesian optimization exclusively for CGYRO identified MTM modes using the same (ne=3, nu=7) velocity resolution. Fig.~\ref{fig:eigs_optim_ne3_nu7_mtm} shows that MTM-specific optimization yields improved performance with parameters ($wz=1.443, nz=32$), achieving RMS errors of 11.42\% (growth rate) and 32.44\% (frequency) with $\sigma_{\rm branch} = 0.135$. Notably, the MTM-optimal width $wz=1.443$ is larger than the mixed-mode optimal value of $wz=0.83$.

\begin{figure}[h!]
    \centering    
        \includegraphics[width=0.99\textwidth]{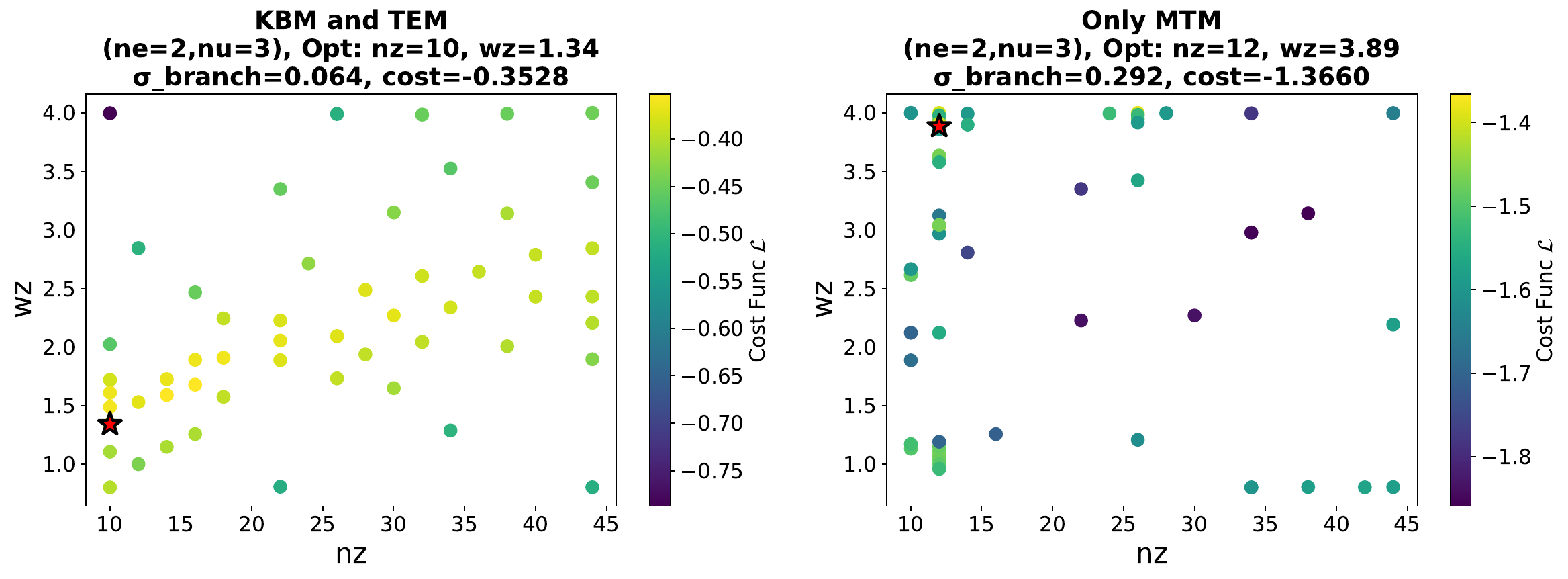}
    \caption{Comparison of Bayesian optimization results for $(ne=2, nu=3)$ configuration, showing separate optimizations for KBM/TEM modes versus MTM modes. The different optimal $wz$ values highlight the challenge of finding universally optimal parameters.}
    \label{fig:GFSne2nu3_optimization}
\end{figure}

These differing optimal spatial parameters between mode types also help explain the poor frequency performance observed for the lower velocity resolution ($ne=2, nu=3$) configuration. To investigate this further, we performed separate Bayesian optimizations for (KBM \& TEM) and MTM modes at low velocity resolution. Fig.~\ref{fig:GFSne2nu3_optimization} reveals that these mode types require substantially different $wz$ values: KBM/TEM modes optimize at ($wz=1.34, nz=10$) while MTMs prefer ($wz=3.89, nz=12$). This divergence makes it challenging to find universally optimal parameters for mixed-mode databases using the Hermite basis for the along the field line coordinate.

\begin{figure}[h!]
    \centering    
        \includegraphics[width=0.99\textwidth]{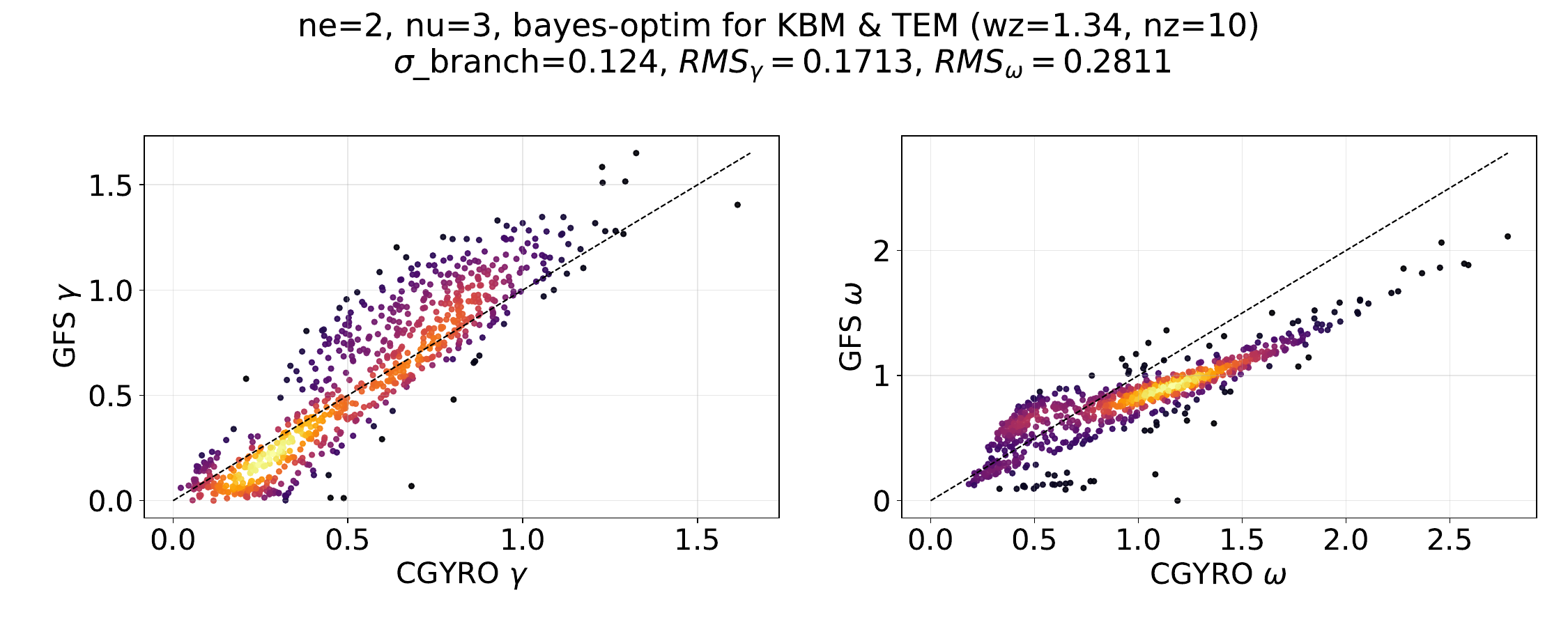}
        \includegraphics[width=0.99\textwidth]{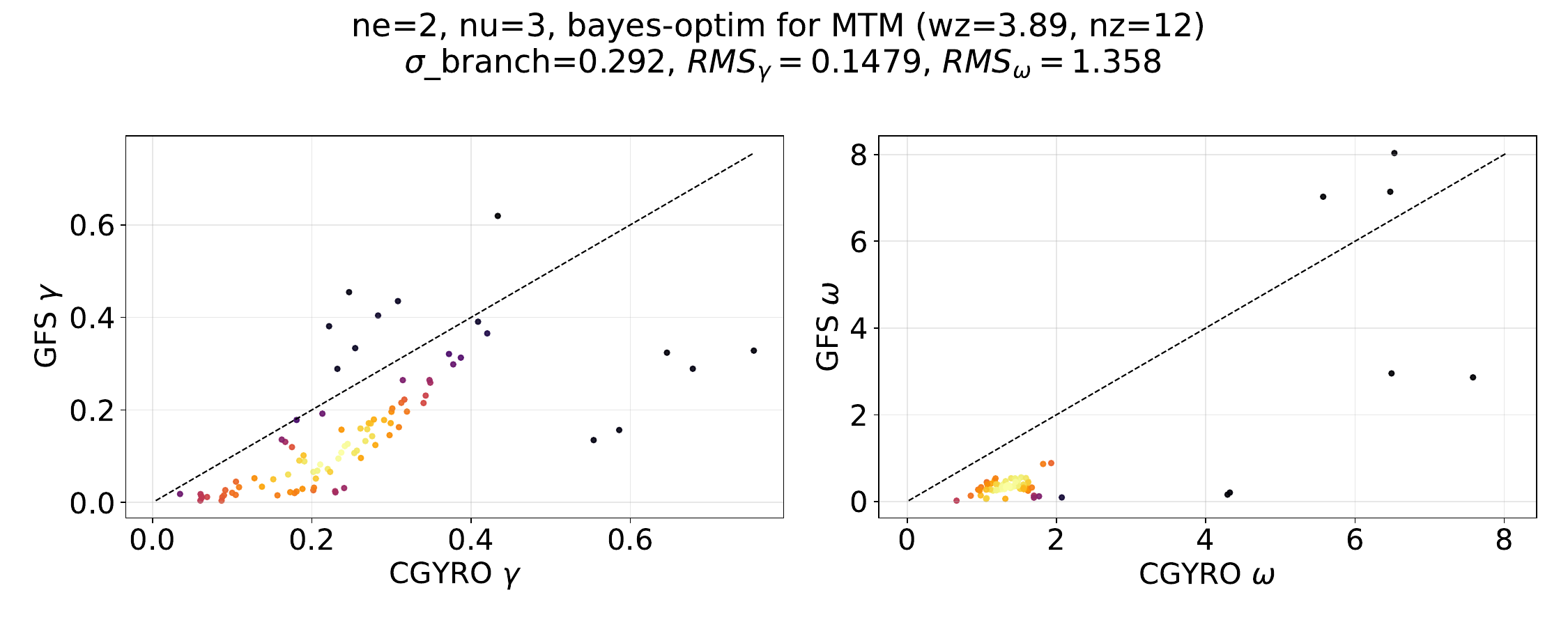}
    \caption{Mode-specific eigenvalue comparisons for $(ne=2, nu=3)$ configuration using mode-optimized parameters. Top: KBM and TEM modes show good agreement with CGYRO using $(wz=1.34, nz=10)$. Bottom: MTM modes exhibit larger errors using $(wz=3.89, nz=12)$, indicating the need for higher velocity resolution.}
    \label{fig:eigGFSne2nu3}
\end{figure}

Fig.~\ref{fig:eigGFSne2nu3} demonstrates the consequences of this parameter divergence. When using mode-specific optimal parameters, GFS with low velocity resolution ($ne=2, nu=3$) can accurately simulate KBM and TEM eigenvalues, achieving RMS errors of 17.13\% (growth rate) and 28.11\% (frequency) with good mode identification ($\sigma_{\rm branch} = 0.124$). However, the same low velocity resolution produces much larger frequency errors for MTM modes (RMS error = 135.8\%) despite reasonable growth rate accuracy (14.79\%), along with poor mode identification ($\sigma_{\rm branch} = 0.292$). It indicates that GFS with low velocity resolution fails to simulate eigenvalues, especially mode frequency, for MTMs.

\section{Comparison with TGLF results}\label{sec:tglf}

To conclude the simulation test, we benchmark CGYRO against TGLF (SAT2 version of collision model \cite{Staebler:2021}) using standard Miller geometry to provide context for the GFS performance demonstrated in the previous sections. To ensure fair comparison, we apply the same Bayesian optimization methodology to TGLF as used for GFS. The optimization targets the same spatial resolution parameters $(wz, nz)$ within the parameter space $([10,44], [0.8,4.0])$, using identical cost function formulation and search procedures.

\begin{figure}[h!]
    \centering    
        \includegraphics[width=0.5\textwidth]{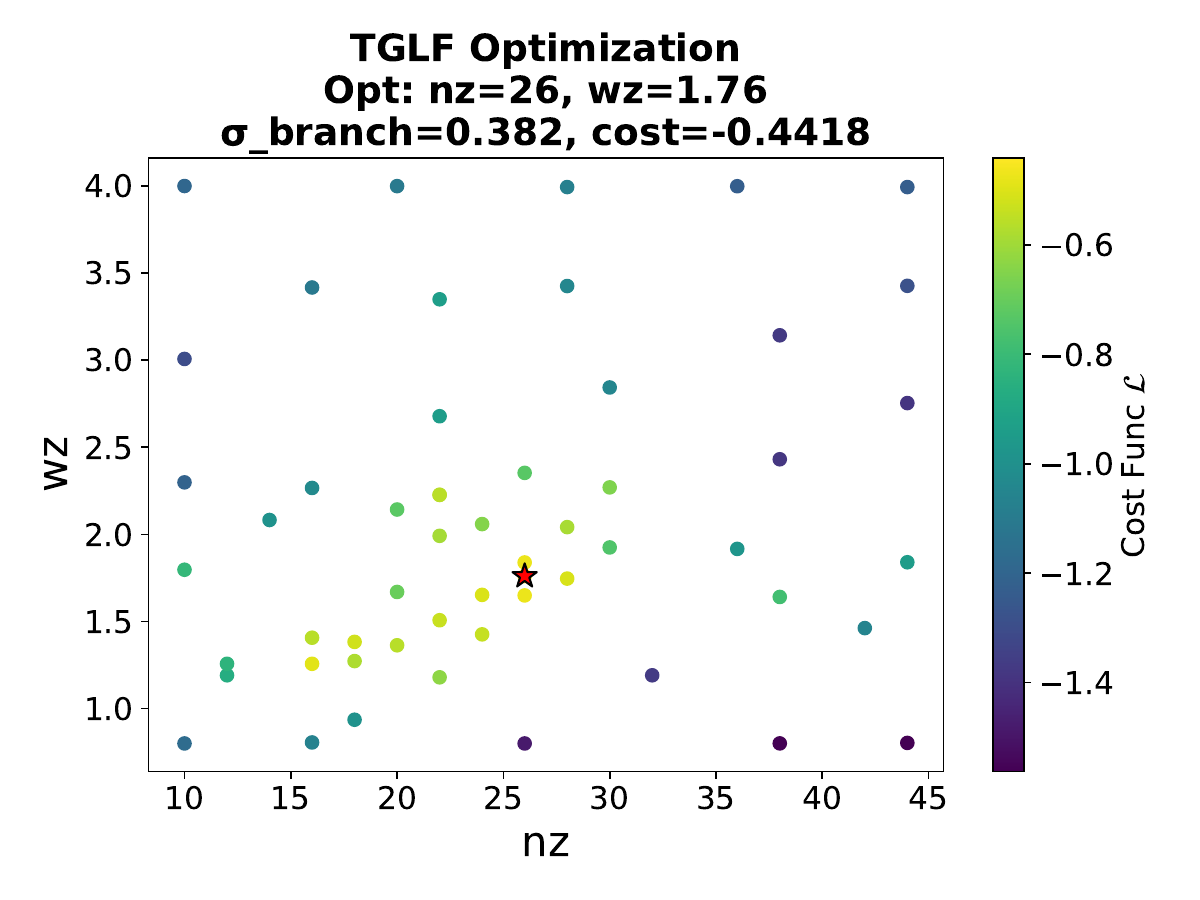}
    \caption{Bayesian optimization results for TGLF showing scatter plot of $(nz, wz)$ colored by cost function $\mathcal{L}$. The red star indicates the optimal point from 42 search evaluations, achieving maximum $\mathcal{L} = -0.4418$ at $(nz,\,wz) = (26,\,1.76)$.}
    \label{fig:tglf_bayes}
\end{figure}
Fig.~\ref{fig:tglf_bayes} shows the TGLF optimization results. Unlike the GFS optimization landscape, which exhibited multiple local optima, the TGLF optimization displays a smoother cost function surface with a more clearly defined global optimum at $(wz=1.76, nz=26)$ with $\mathcal{L}=-0.4418$. This smoother optimization landscape suggests that TGLF may be less sensitive to parameter variations within the tested range, though this comes at the cost of overall performance.

\begin{figure}[h!]
    \centering    
    \includegraphics[width=0.99\textwidth]{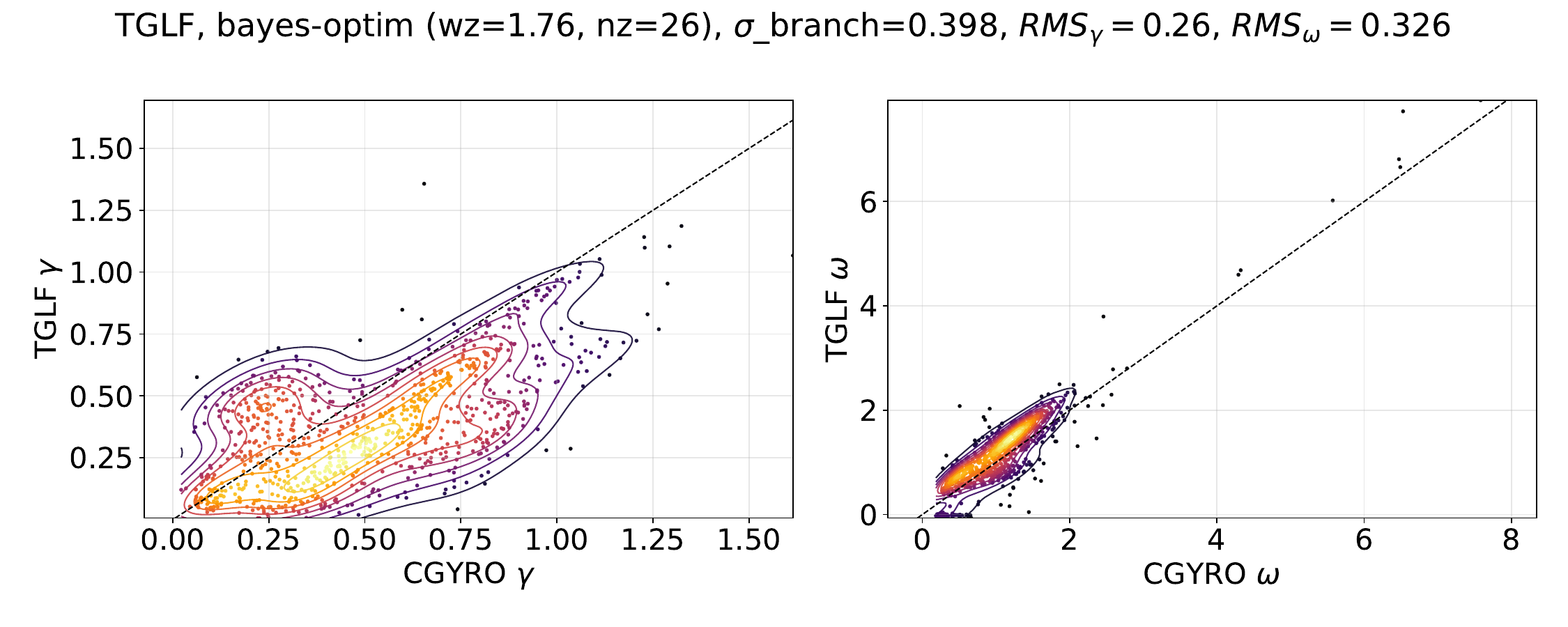}    \caption{Eigenvalue comparison between CGYRO (x-axis) and TGLF (y-axis), showing RMS error of growth rate $\gamma \sim 26.00\%$ (left) and RMS error of frequency $\sim 32.60\%$ (right). TGLF has an unfavorable $\sigma_{\rm branch}=0.398$.}
    \label{fig:eigs_tglf}
\end{figure}

Fig.~\ref{fig:eigs_tglf} presents the eigenvalue comparison between CGYRO and TGLF with optimized $(wz,nz)$ across the complete NSTX database. TGLF achieves RMS errors of 26.00\% for growth rate and 32.60\% for frequency. However, TGLF exhibits a significant mode identification challenge with $\sigma_{\rm branch} = 39.8\%$, indicating that nearly 40\% of cases require selection of subdominant modes to match CGYRO results.

Table.~\ref{tab:performance_comparison} summarizes the comparative performance of optimized GFS $(ne=3, nu=7)$ and TGLF against CGYRO for NSTX pedestal conditions. The results demonstrate significant advantages of the optimized GFS approach across all metrics.

\begin{table}[h!]
\centering
\caption{Performance comparison of GFS and TGLF against CGYRO}
\label{tab:performance_comparison}
{\small\begin{tabular}{l c c c c}
\toprule
\textbf{Configuration} &  \textbf{Growth Rate} & \textbf{Frequency} & \textbf{Mode Mismatch} \\
&  \textbf{RMS Error (\%)} & \textbf{RMS Error (\%)} & $\sigma_{\rm branch}$ (\%) \\
\midrule
GFS $(ne=3, nu=7)$& $15.91$ & $21.11$ & $10$ \\
TGLF & $26.00$ & $32.60$ & $39.8$ \\
\bottomrule
\end{tabular}}
\end{table}

\section{Analysis of parameter dependence of eigenvalue errors}\label{sec:analyerror}

This section examines the performance of GFS by analyzing eigenvalue errors as functions of key plasma parameters: minor radius $r/a$, magnetic shear $ \hat{s}=(r/q)d q/dr$, and poloidal wavenumber $k_y$. This investigation aims to identify where the large discrepancies and mode identification differences between GFS and CGYRO occur. By examining error distributions across different plasma parameter ranges, we can determine the specific conditions under which GFS struggles to reproduce CGYRO results and understand the underlying causes of these disagreements. 
%Normalized pedestal radius $\rho_{\rm ped}$ in our simulation is a dimensionless radial index defined by dividing the pedestal region—from pedestal bottom to pedestal top—into 13 equally spaced flux surfaces in poloidal flux. The index ranges from 0 to 12, with each value representing a specific flux surface. The central surface ($\rho_{\rm ped}=6/12$) corresponds to the region with the steepest gradient. 

We analyze the full dataset of 864 simulation cases, encompassing both KBM, TEM and MTM scenarios, by categorizing the simulations based on $r/a$ and $k_y$. The Root Mean Square (RMS) errors are used to quantify performances in growth rates and frequencies between CGYRO and GFS for two optimal resolution settings: $(ne=2, nu=3, nz=20, wz=3.80)$ and $(ne=3, nu=7, nz=34, wz=0.83)$.

\subsection{Eigenvalue errors as functions of radius}

\begin{figure}[h!]
    \centering    
    \includegraphics[width=0.99\textwidth]{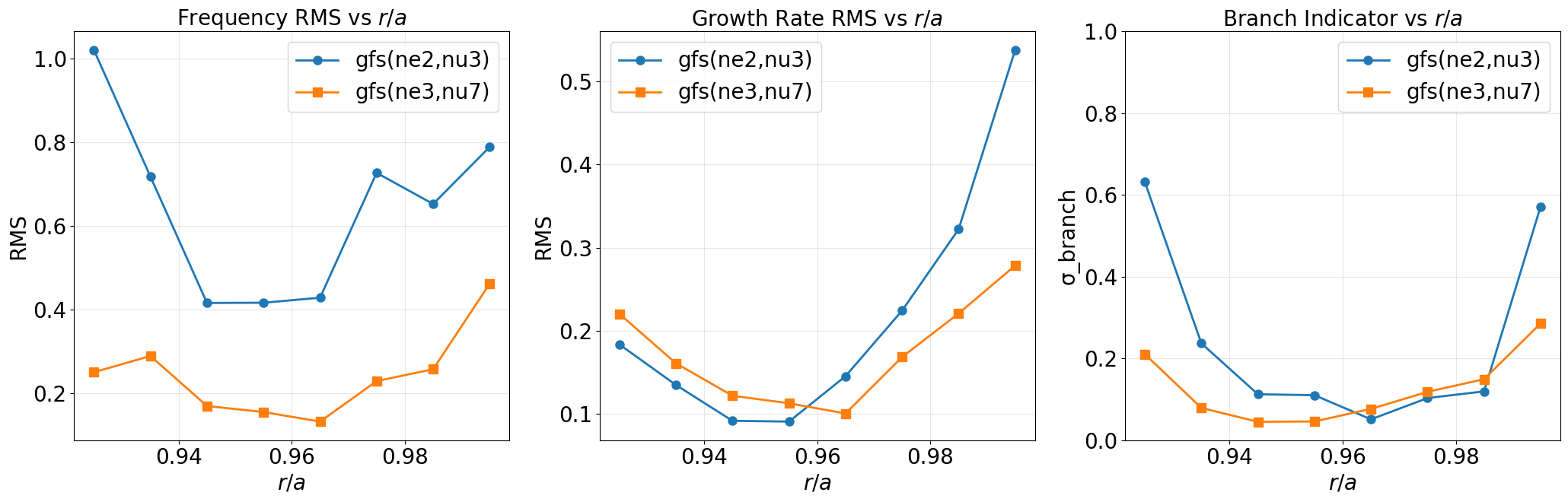}
    \caption{RMS errors in frequency (left), growth rate (middle), and mode mismatch fraction $\sigma_{\rm branch}$ (right) between CGYRO and GFS as a function of minor radius $r/a$. The results compare two resolution settings: $(ne=2, nu=3)$ with optimal $(nz=20, wz=3.80)$ and $(ne=3, nu=7)$ with optimal $(nz=34, wz=0.83)$. Both configurations show increasing errors toward the separatrix, with the low-resolution setting exhibiting significantly larger frequency errors in the inner pedestal region.}
    \label{rms_rho}
\end{figure}

Fig.~\ref{rms_rho} presents the RMS error in mode frequency (left) and growth rate (middle) and mode mismatch fraction $\sigma_{\rm branch}$ (right) between CGYRO and GFS as a function of the normalized minor radius $r/a$ in the pedestal region. The analysis is performed for two GFS velocity-space resolution settings $(ne=2, nu=3)$ and $(ne=3, nu=7)$ with their optimal ($nz, wz$). The radial domain in the pedestal region spans $r/a \in [0.92, 1.0]$, where $r/a=1$ corresponds to the separatrix. Each bin represents a discrete subregion within the pedestal, spaced uniformly in poloidal flux.

For frequency RMS errors (left), the two resolution settings exhibit markedly different behaviors. The low-resolution setting $(ne=2, nu=3)$ (blue) shows large frequency errors ($\sim 1.0$) at the innermost radial location ($r/a \sim 0.92-0.93$), which then decrease significantly and remain relatively stable at $\sim 0.4-0.7$ across the middle pedestal region, before showing some increase toward the separatrix. In contrast, the high-resolution setting $(ne=3, nu=7)$ (orange) demonstrates much better frequency accuracy in the inner and middle pedestal regions, with RMS errors remaining low ($\sim 0.15-0.3$) across most radial locations, but exhibits a notable increase to $\sim 0.45$ only in the outermost region near the separatrix.

Growth rate RMS errors (middle) show relatively similar patterns for both resolution settings, remaining low ($\sim 0.1-0.2$) across most of the pedestal but exhibiting significant increases near the separatrix. The $(ne=2, nu=3)$ configuration shows growth rate errors increasing dramatically to $\sim 0.55$ at the separatrix, while the $(ne=3, nu=7)$ configuration shows a more moderate increase to $\sim 0.28$. Both configurations demonstrate that growth rate prediction becomes more challenging in the steep gradient region closest to the plasma edge.

The mode mismatch fraction $\sigma_{\rm branch}$ (right) reveals important insights into mode identification challenges. The $(ne=2, nu=3)$ configuration exhibits high mode mismatch rates ($\sim 0.6$) in the innermost pedestal region, which decrease to $\sim 0.2-0.25$ in the middle pedestal before increasing again toward the separatrix ($\sim 0.55$). The $(ne=3, nu=7)$ configuration shows significantly better mode identification performance, with mismatch rates remaining low ($\sim 0.05-0.15$) across most of the pedestal and only increasing to $\sim 0.3$ near the separatrix. These results indicate that the most significant disagreements between GFS and CGYRO, particularly the mode identification issues, are concentrated in the innermost pedestal region for low-resolution settings and near the plasma edge for both configurations, where the codes face challenges in accurately resolving the complex physics.

\subsection{Eigenvalue errors as functions of magnetic shear}

\begin{figure}[h!]
    \centering
    \includegraphics[width=0.99\textwidth]{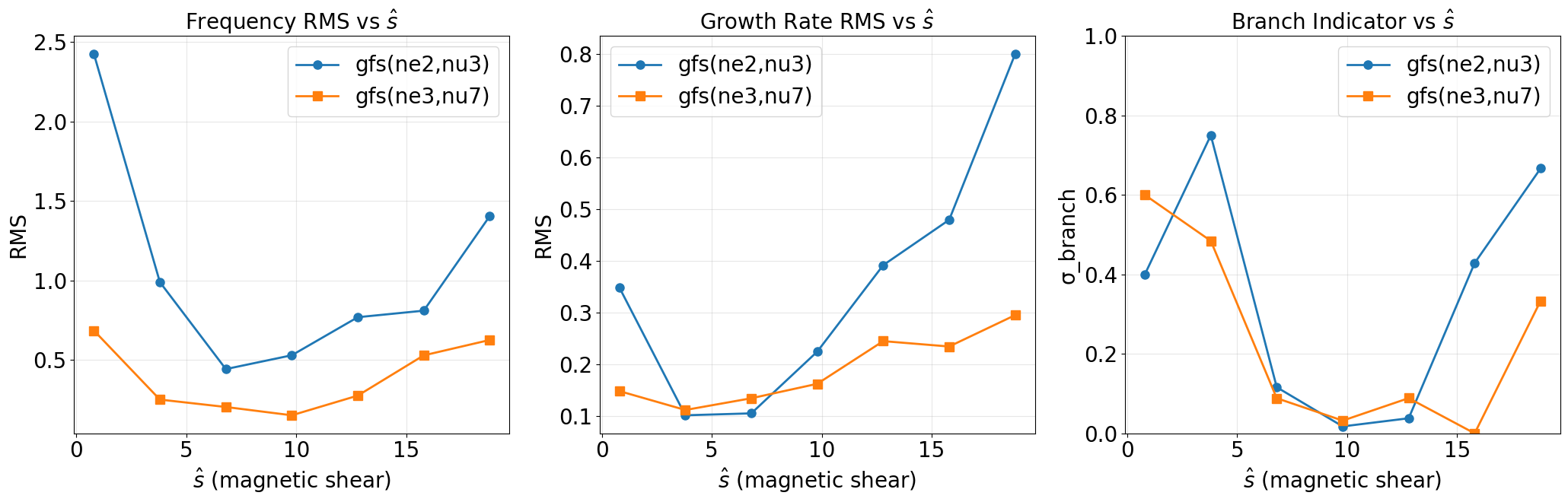}
    \caption{RMS errors in frequency (left), growth rate (middle), and mode mismatch fraction $\sigma_{\rm branch}$ (right) between CGYRO and GFS as a function of magnetic shear $\hat{s}$. Results compare two GFS velocity-space resolution settings: $(ne=2, nu=3)$ with optimal $(nz=20, wz=3.80)$ and $(ne=3, nu=7)$ with optimal $(nz=34, wz=0.83)$. The dramatic frequency error peaks at low $\hat{s}$ values indicate mode identification issues in low magnetic shear regions of the pedestal.}
    \label{fig:rms_qprime}
\end{figure}

Fig.~\ref{fig:rms_qprime} presents the RMS error in mode frequency (left) and growth rate (middle) and mode mismatch fraction $\sigma_{\rm branch}$ (right) between CGYRO and GFS as a function of the magnetic shear $\hat{s}$ at the mode location. Each bin groups cases with similar $\hat{s}$ values to examine the dependence of eigenvalue discrepancies on local magnetic shear conditions.

The frequency RMS errors reveal distinct patterns for the two resolution settings. For $(ne=2, nu=3)$ (blue), frequency errors show a dramatic peak at the lowest $\hat{s}$ values ($\hat{s} \sim 0-2$), reaching approximately 2.4, before decreasing significantly and remaining relatively stable at $\sim 0.4-1.4$ across the higher $\hat{s}$ range. In contrast, the $(ne=3, nu=7)$ (orange) configuration exhibits much better frequency accuracy across most $\hat{s}$ values, with RMS errors remaining low ($\sim 0.1-0.3$) for intermediate $\hat{s}$ ranges, but shows a notable peak reaching $\sim 0.7$ at the lowest $\hat{s}$ values and some increase to $\sim 0.6$ at the highest $\hat{s}$ values. This pronounced peak at low $\hat{s}$ for both configurations indicates that mode identification mismatches are predominantly occurring in low magnetic shear regions.

Growth rate RMS errors display different behavior compared to frequency errors. The $(ne=2, nu=3)$ configuration shows growth rate errors starting at $\sim 0.35$ for low $\hat{s}$, decreasing to $\sim 0.1$ for intermediate $\hat{s}$ values, then increasing dramatically to $\sim 0.8$ at the highest $\hat{s}$ values. The $(ne=3, nu=7)$ setting exhibits more uniform and lower growth rate errors, remaining around $0.1-0.3$ across most of the $\hat{s}$ range, with only a moderate increase to $\sim 0.3$ at the highest $\hat{s}$ values.

The mode mismatch fraction $\sigma_{\rm branch}$ reveals important correlations with magnetic shear. The $(ne=2, nu=3)$ configuration shows high mode mismatch rates ($\sim 0.4-0.75$) at both low and high $\hat{s}$ extremes, with better mode identification ($\sim 0.1$) in the intermediate $\hat{s}$ range. The $(ne=3, nu=7)$ configuration demonstrates significantly better overall mode identification performance, with mismatch rates remaining low ($\sim 0.05-0.1$) across most $\hat{s}$ values, showing increases only at the lowest ($\sim 0.6$) and highest ($\sim 0.35$) $\hat{s}$ values.

\subsection{Two-dimensional error distribution: combined radius and magnetic shear effects}

\begin{figure}[h!]
    \centering    
    \includegraphics[width=0.99\textwidth]{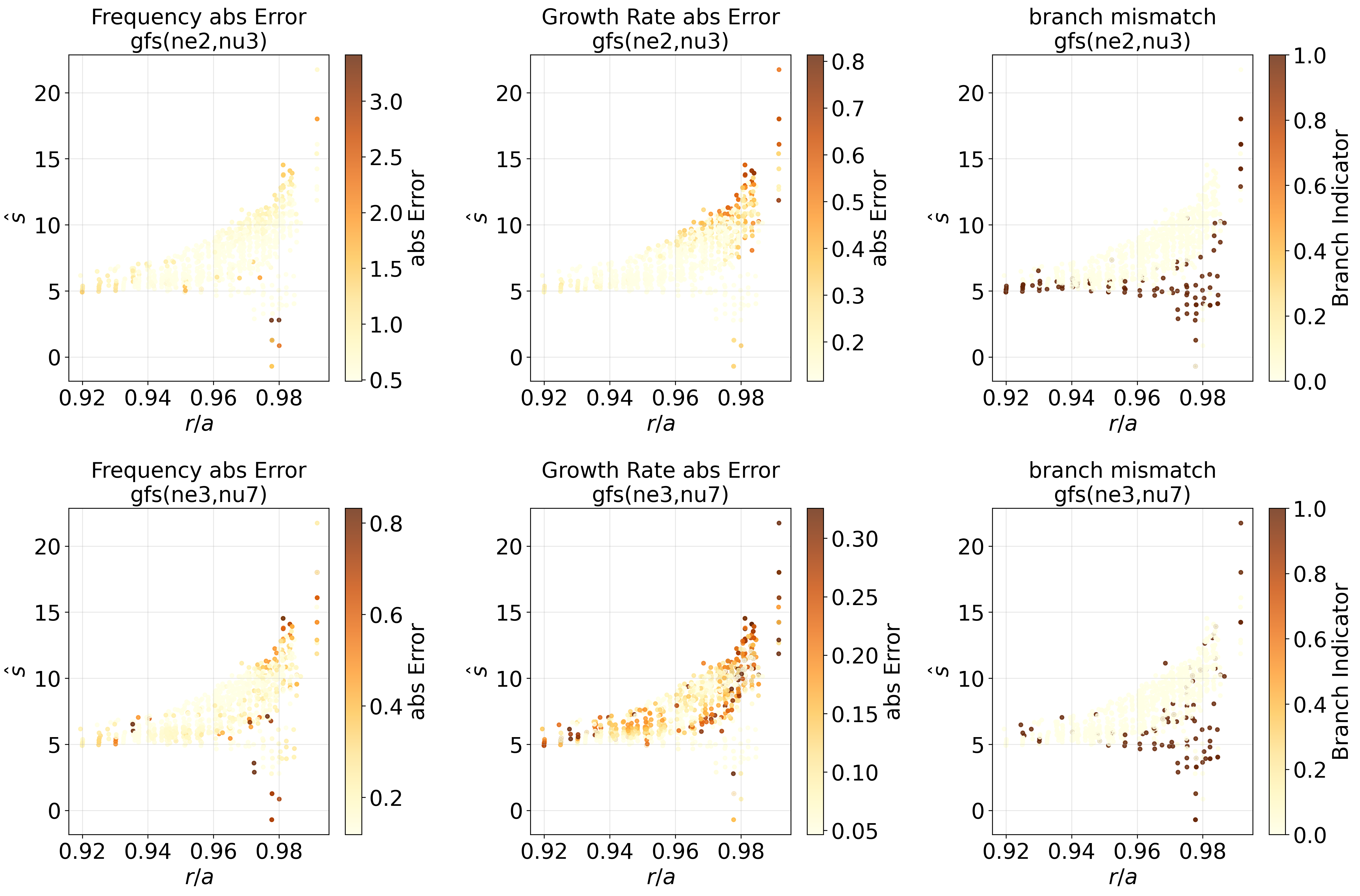}
\caption{Two-dimensional scatter plots showing absolute errors in frequency (left), growth rate (middle), and mode mismatch (right) as functions of minor radius $r/a$ and magnetic shear $\hat{s}$. Top row shows $(ne=2, nu=3)$ configuration, bottom row shows $(ne=3, nu=7)$ configuration. Color intensity indicates error magnitude, with darker colors representing larger errors. Large errors are concentrated in low magnetic shear regions and near the plasma edge. Note that colormap scales differ between configurations due to different error magnitudes.}
    \label{fig:2d_scatter}
\end{figure}

To examine potential correlations between the radial and magnetic shear dependencies observed in the previous sections, Fig~\ref{fig:2d_scatter} presents two-dimensional scatter plots showing absolute errors in frequency, growth rate, and mode mismatch as functions of both minor radius $r/a$ and magnetic shear $\hat{s}$.

The analysis reveals that large frequency errors are concentrated in two distinct parameter regions: low magnetic shear ($\hat{s} < 5$) across all radial locations, and the plasma edge region ($r/a > 0.97$) regardless of magnetic shear. This demonstrates that the error dependencies on $r/a$ and $\hat{s}$ are largely independent rather than correlated. Growth rate errors follow a similar pattern, with the highest errors occurring primarily in the high radius, high magnetic shear region ($r/a > 0.97$, $\hat{s} > 10$).

Mode identification mismatches also occur in these same two parameter regions: the low $\hat{s}$ domain and the high $r/a$ region. This confirms that mode identification failures result from specific combinations of challenging parameter values rather than a single underlying correlation.

The majority of simulation points lie in the favorable parameter space of moderate radius ($0.93 < r/a < 0.97$) and moderate to high magnetic shear ($\hat{s} > 5$), explaining why overall RMS errors appear moderate despite substantial individual case discrepancies in the challenging parameter regions.

\subsection{Eigenvalue errors as functions of poloidal wavenumber}

\begin{figure}[h!]
    \centering    
    \includegraphics[width=0.99\textwidth]{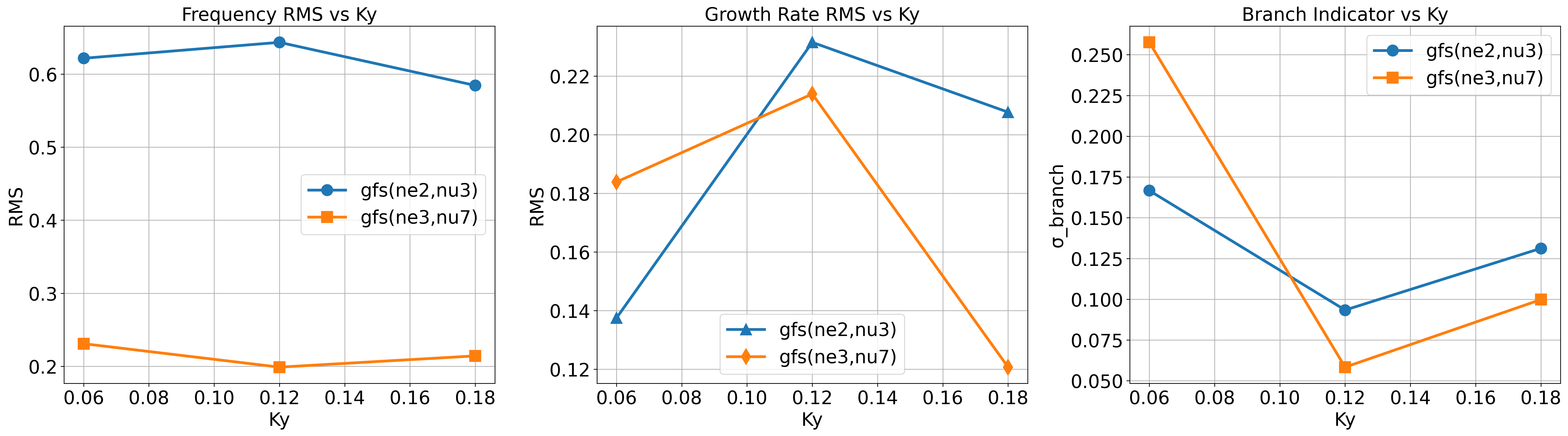}
    \caption{RMS errors in frequency (left), growth rate (middle), and mode mismatch fraction $\sigma_{\rm branch}$ (right) between CGYRO and GFS as a function of normalized poloidal wavenumber $k_y$. Results compare $(ne=2, nu=3)$ and $(ne=3, nu=7)$ configurations. Both configurations show similar patterns with peaks in growth rate errors and minima in mode mismatch at intermediate $k_y$ values.}
    \label{fig:rms_ky}
\end{figure}

Fig.~\ref{fig:rms_ky} presents the RMS error in mode frequency (left) and growth rate (middle) and mode mismatch fraction $\sigma_{\rm branch}$ (right) between CGYRO and GFS as a function of the normalized poloidal wavenumber $k_y$.

Both velocity resolution configurations exhibit similar patterns across the wavenumber range. For frequency errors, the low-resolution $(ne=2, nu=3)$ configuration consistently shows larger errors (around 0.6) compared to the high-resolution $(ne=3, nu=7)$ configuration (around 0.2), with both maintaining relatively stable error levels across wavenumbers.

Growth rate errors show a characteristic peak at intermediate wavenumbers ($k_y \sim 0.12$) for both configurations. Mode identification performance shows some variation with wavenumber, with $\sigma_{\rm branch}$ reaching minimum values around $k_y = 0.12-0.14$ for both configurations.

\section{Conclusion}\label{sec:conclusion} 

This study validates the Gyro-Fluid System (GFS) model for analyzing pedestal microinstabilities in NSTX spherical tokamak plasmas through systematic benchmarking against CGYRO and TGLF. We introduce a novel Bayesian optimization approach to identify optimal resolution parameters specifically tailored for spherical tokamak pedestal conditions.

The optimized high-resolution configuration $(ne = 3, nu = 7, wz = 0.83, nz = 34)$ achieves exceptional accuracy with RMS errors of 15.91\% (growth rate) and 21.11\% (frequency) with only 10\% mode identification mismatch. Mode-specific analysis reveals that GFS excels at capturing kinetic ballooning mode (KBM) and trapped electron mode (TEM) physics, achieving RMS errors below 20\% for both growth rates and frequencies with excellent mode identification ($\sigma_{branch} < 10\%$).
Micro-tearing mode (MTM) simulation presents greater eigenvalue errors, requiring different optimal spatial parameters with wider $wz$ based on simulation results. This mode-dependent optimization requirement indicates that different instability types benefit from specialized parameter tuning.

The lower resolution configuration $(ne = 2, nu = 3)$ achieves competitive accuracy for KBM and TEM modes but fails for MTM modes, indicating that MTMs require higher velocity resolution for accurate simulation.

Parametric analysis reveals that eigenvalue discrepancies concentrate in two distinct regions: low magnetic shear conditions ($\hat{s} < 5$) and near the separatrix $(r/a > 0.97)$ where extreme gradients and electromagnetic effects degrade predictions. 

Compared to optimized TGLF's 26.00\% growth rate and 32.60\% frequency RMS errors with 39.8\% mode identification mismatch, optimized GFS demonstrates substantial improvements across all metrics. These results establish GFS as an effective tool for spherical tokamak pedestal stability analysis, providing the kinetic ballooning mode constraints necessary for EPED framework implementation while maintaining computational efficiency suitable for predictive transport modeling.

\section{Acknowledgments}
This material is based upon work supported by the U.S. Department of Energy, Office of Science, Office of Fusion Energy Sciences, Theory Program and the NSTX-U Research Program, using the NSTX-U Fusion Facility, a DOE Office of Science user facility. We thank the NSTX-U experimental team for providing the data analyses.

DISCLAIMER: This report was prepared as an account of work sponsored by an agency of the United States Government. Neither the United States Government nor any agency thereof, nor any of their employees, makes any warranty, express or implied, or assumes any legal liability or responsibility for the accuracy, completeness, or usefulness of any information, apparatus, product, or process disclosed, or represents that its use would not infringe privately owned rights. Reference herein to any specific commercial product, process, or service by trade name, trademark, manufacturer, or otherwise, does not necessarily constitute or imply its endorsement, recommendation, or favoring by the United States Government or any agency thereof. The views and opinions of authors expressed herein do not necessarily state or reflect those of the United States Government or any agency thereof.

Notice: This manuscript has been authored by UT-Battelle, LLC, under contract DE-AC05-00OR22725 with the US Department of Energy (DOE). The US government retains and the publisher, by accepting the article for publication, acknowledges that the US government retains a nonexclusive, paid-up, irrevocable, worldwide license to publish or reproduce the published form of this manuscript, or allow others to do so, for US government purposes. DOE will provide public access to these results of federally sponsored research in accordance with the DOE Public Access Plan (https://www.energy.gov/doe-public-access-plan).

\bibliographystyle{abbrvnat}  % or another style
\bibliography{refs}
\end{document}